\documentclass{ectj}

\usepackage{amsfonts,amssymb,graphics,epsfig,verbatim,bm,latexsym,amsmath,url,amsbsy}
\usepackage[top=3cm, bottom=3cm, left=3cm, right=3cm]{geometry}
\usepackage[english]{babel}
\usepackage{bbm}
\usepackage{booktabs}
\usepackage{mathtools}
\usepackage{setspace}
\usepackage{multirow, multicol}
\usepackage{floatrow}
\usepackage{chbibref}
\usepackage{float}
\usepackage{color}
\usepackage[toc,title]{appendix}
\usepackage{csquotes}
\usepackage{hyperref}
\usepackage{enumerate}
\usepackage{graphicx}
\usepackage{subcaption}
\usepackage{pdflscape}
\usepackage{placeins}
\usepackage{IEEEtrantools}
\usepackage{algorithm}
\usepackage{pstricks}
\usepackage{tikz}
\usetikzlibrary{automata, arrows, positioning, calc}
\usetikzlibrary{positioning}
\newrgbcolor{dgreen}{0 0.5 0}
\hypersetup{
    colorlinks,
    citecolor=blue,
    filecolor=black,
    linkcolor=blue,
    urlcolor=black
}

\newcommand{\RR}{\mathbb{R}}

\newcommand{\EE}{\mathbb{E}}
\newcommand{\NN}{\mathbb{N}}

\newcommand{\simiid}{\overset{\textrm{i.i.d.}}{\sim}}

\newcommand{\Var}{\text{Var}}

\newcommand\independent{\protect\mathpalette{\protect\independenT}{\perp}}

\DeclareMathOperator*{\argmin}{arg\,min}
\providecommand{\keywords}[1]{\textbf{\textit{Keywords:}} #1}
\makeatletter
\newcommand*{\centernot}{%
  \mathpalette\@centernot
}
\def\@centernot#1#2{%
  \mathrel{%
    \rlap{%
      \settowidth\dimen@{$\m@th#1{#2}$}%
      \kern.5\dimen@
      \settowidth\dimen@{$\m@th#1=$}%
      \kern-.5\dimen@
      $\m@th#1\not$%
    }%
    {#2}%
  }%
}
\makeatother
\def\independenT#1#2{\mathrel{\rlap{$#1#2$}\mkern2mu{#1#2}}}
\newcommand{\notindependent}{\centernot{\independent}}
\newcommand{\plim}{\operatorname*{plim}}

\newtheorem{theorem}{Theorem}
\newtheorem{assumption}{Assumption}
\newtheorem{proposition}{Proposition}

\newtheorem{lemma}{Lemma}

\newtheorem{remark}{Remark}

\renewcommand{\thesection}{\arabic{section}}
\renewcommand{\theequation}{\arabic{section}.\arabic{equation}}

\title[Team Networks with Partially Observed Links]{Team Networks with Partially Observed Links}

\author[Y. Xu]{Yang~Xu$^{\dagger}$}
\address{$^{\dagger}$ Department of Economics and Nuffield College, University of Oxford.}
\email{yang.xu@economics.ox.ac.uk}

\def\AmSTeX{$\cal A$\kern-.1667em\lower.5ex\hbox{$\cal M$}\kern-.125em
    $\cal S$-\TeX}
\def\BibTeX{{\rm B\kern-.05em{\sc i\kern-.025em b}\kern-.08em
    T\kern-.1667em\lower.7ex\hbox{E}\kern-.125emX}}

\begin{document}
\let\enquote\relax
\begin{abstract}
    This paper studies a linear production model in team networks with missing links. In the model, heterogeneous workers, represented as nodes, produce jointly and repeatedly within teams, represented as links. Links are omitted when their associated outcome variables fall below a threshold, resulting in partial observability of the network. To address this, I propose a Generalized Method of Moments estimator under normally distributed errors and develop a distribution-free test for detecting link truncation. Applied to academic publication data, the estimator reveals and corrects a substantial downward bias in the estimated scaling factor that aggregates individual fixed effects into team-specific fixed effects. This finding suggests that the collaboration premium may be systematically underestimated when missing links are not properly accounted for.
\end{abstract}

\section{Introduction}
In economic networks, nodes or agents generally have individual-specific heterogeneity that can be modeled through a fixed-effect approach. At the same time, links or edges also exhibit unobserved heterogeneity, which may relate to the node-level heterogeneity. In settings such as team production, networks provide a structured approach for analyzing complex relational dependencies between heterogeneous workers and heterogeneous teams (\citet{bonhomme_teams_2021}). However, real-world networks are often partially observed, and the standard assumption of random sampling may sometimes fail in network models (\citet{chandrasekhar_econometrics_2011}). In particular, partial network observability implies that only a selected subset of links and nodes is observed, while others are systematically omitted. Therefore, new econometric tools may be needed to correct for the selection bias inherent in such endogenous network. This paper proposes a truncation-robust estimator in a linear production model under normally distributed errors, highlighting the importance of distinguishing between the observed network and the latent network.

In a team network, nodes represent workers, and links represent teams. The network can be modeled as a hypergraph—that is, a graph whose links can join any number of nodes. To capture repeated teamwork within a team, multiple links may exist among a fixed set of nodes, resulting in a multigraph representation. The minimum team size is one, with self-loops representing solo production. Economic production occurs at the link level $\ell$, with an associated latent outcome variable $Y_\ell^*$. In an ideal world, one would fully observe the latent network, which consists of the latent sets of links and nodes. However, links with negative latent outcome $Y_{\ell}^* < 0$ are truncated. Therefore, only the truncated graph is observed, consisting of a subset of links and nodes. In this setting, node missingness occurs when all of a node's links are omitted, constituting an extreme case of link truncation. This differs from censoring: if a latent outcome $Y_{\ell}^*$ falls below a censoring threshold, it is replaced with a censored value (e.g., zero dollars in the case of censored wage), but its team composition remains observed.

The partially observed network model nests the fully observed model, which assumes no missing links. A stylized example of a network before and after link truncation is shown below. In Figure \ref{fig:stylized_1a}, half of the links have non-negative latent outcomes (represented by solid lines), while the other half have negative outcomes (represented by dashed lines). After link truncation, the observed graph only contains three links and three nodes, as shown in Figure \ref{fig:stylized_1b}. Note that node $D$ (represented as a dashed circle) is missing because its only joint link with $A$ is omitted. In this setting, the truncated links are entirely omitted, rendering both the outcome variable and the team composition unobserved. The proposed estimator is appealing, as it relies only on the partially observed network and flexibly accommodates missing links and nodes.

\begin{figure}[H]
    \centering
    {\begin{subfigure}{0.45\textwidth}
    \centering
        \tikzset{auto shift/.style={auto=right, -, to path={ let \p1=(\tikztostart),\p2=(\tikztotarget), \n1={atan2(\y2-\y1,\x2-\x1)},\n2={\n1+180} in ($(\tikztostart.{\n1})!1mm!270:(\tikztotarget.{\n2})$) -- ($(\tikztotarget.{\n2})!1mm!90:(\tikztostart.{\n1})$) \tikztonodes}}}
        \begin{tikzpicture} 
            \node[state](a) {A};
            \node[state, right=of a] (b) {B};  
            \node[state, below=of b] (c) {C};  
            \node[state, dashed, below=of a] (d) {D};
        \draw
        (a) edge[auto shift] [dashed] node {} (b)  
        (b) edge[auto shift] node {} (a);  
        \draw
        (b) edge[auto shift] node {} (c)  
        (c) edge[auto shift] node {} (b);
        \draw
        (a) edge[auto] [dashed] node {} (c);
        \draw
        (a) edge[auto] [dashed] node {} (d);
        \end{tikzpicture}
    \caption{Latent Graph $\mathcal{G}^*$}
    \label{fig:stylized_1a}
    \end{subfigure}}\hfill
    {\begin{subfigure}{0.55\textwidth}
    \centering
        \tikzset{auto shift/.style={auto=right,-,
        to path={ let \p1=(\tikztostart),\p2=(\tikztotarget),
        \n1={atan2(\y2-\y1,\x2-\x1)},\n2={\n1+180}
        in ($(\tikztostart.{\n1})!1mm!270:(\tikztotarget.{\n2})$) -- 
        ($(\tikztotarget.{\n2})!1mm!90:(\tikztostart.{\n1})$) \tikztonodes}}}
        \begin{tikzpicture} 
        \node[state](a) {A};
        \node[state, right=of a] (b) {B};  
        \node[state, below=of b] (c) {C};  
        \draw
        (a) edge[auto] node {} (b);  
        \draw
        (b) edge[auto shift] node {} (c)  
        (c) edge[auto shift] node {} (b);
        \end{tikzpicture}
    \caption{Observed Graph $\mathcal{G}$ with Missing Links and Nodes}
    \label{fig:stylized_1b}
    \end{subfigure}}
    \caption{Stylized Example}
    \label{fig:stylized_example}

    \footnotesize{\textit{Notes: In this stylized example, the latent network has four nodes — A, B, C, and D. Solid lines represent links with non-negative outcome. Dashed lines represent links that have negative outcomes and are hence unobserved. Nodes are observed if at least one of their links is observed. Observed nodes are represented as solid circles, whereas unobserved nodes are represented as dashed circles. The left figure plots a latent network with four nodes and six links. The right figure plots a partially observed network consisting of three nodes and three links, a strict sub-graph of the latent graph.}}
\end{figure}

\begin{figure}[!ht]
    \centering
    \includegraphics[width=0.85\linewidth]{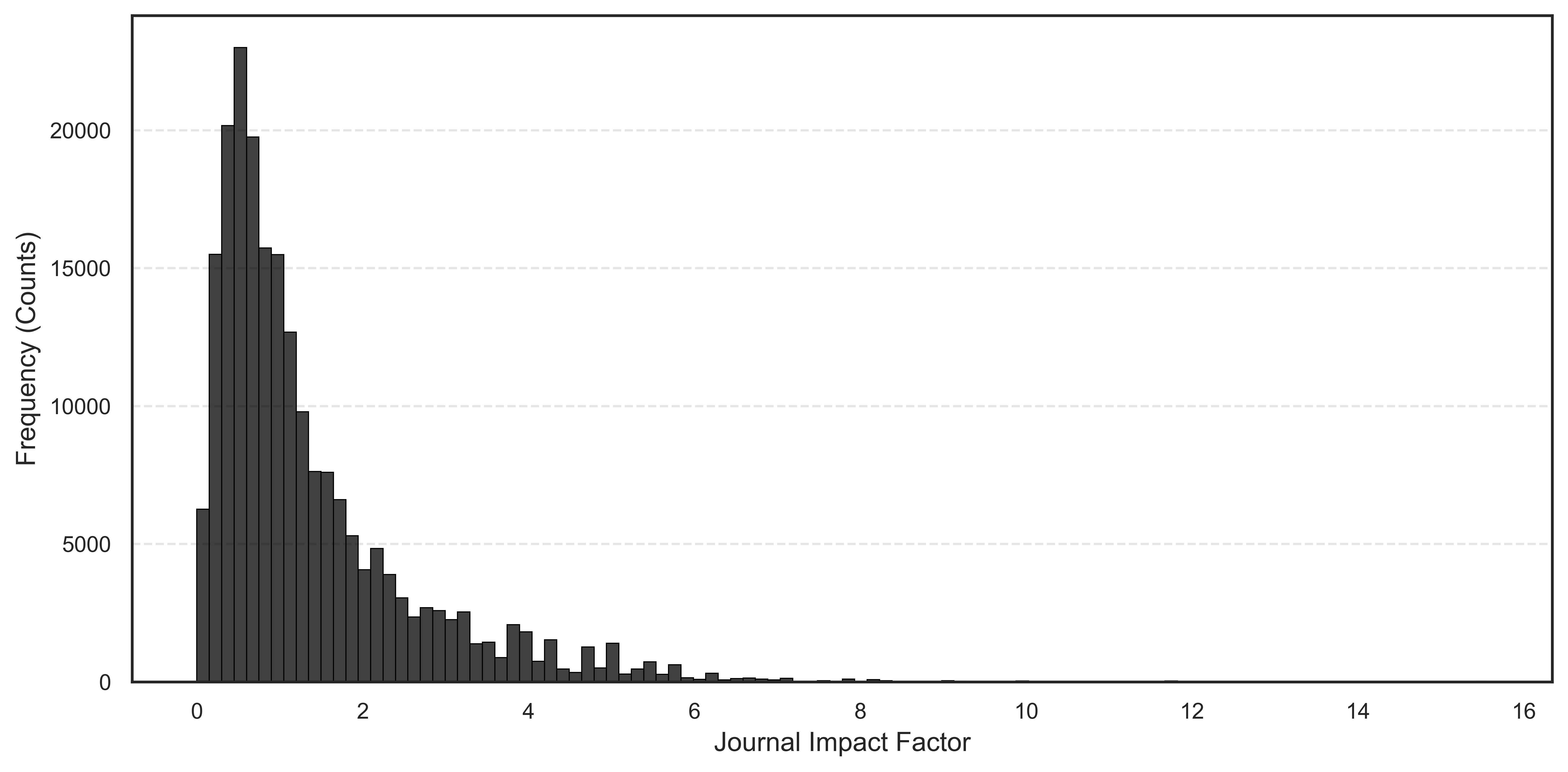}
    \caption{Empirical Distribution of Paper Quality}
    \label{fig:quality_hist}
    \floatfoot{\footnotesize{\textit{Notes: This figure presents the histogram of the paper quality of articles published in economic journals. Paper quality is matched with journal-year impact factor through Clarivate Web of Science Journal Citation Reports (1997 to 2020). The vertical axis measures the article count whereas the horizontal axis measures the impact factor. See Section \ref{sec:application} for details on the data and the sample selection.}}}
\end{figure}

Teams arise naturally in economic production, in which heterogeneous workers collaborate within different teams (\citet{jones_rise_2021}). Examples include patent production (\citet{jaravel_team-specific_2018}; \citet{kerr_global_2018}), software development (\citet{el-komboz_productivity_2024}), factory production \citep{hjort_ethnic_2014}, and professional sports (\citet{devereux_identifying_2018}), among others. In academia, collaboration has increasingly become an integral part of scientific research in many fields spanning from physical sciences to social sciences (\citet{wuchty_increasing_2007}). Researchers work both individually and collaboratively in teams of various sizes. As an innovative production process, research production is almost invariably subject to uncertainty as researchers ask and study novel questions. This means that while some research projects result in publication, others do not bear fruit. As a result, many unpublished projects are systematically omitted from the publication data. Among these unpublished papers, neither the edge weight (paper quality) nor the edge itself (team composition) is observed.\footnote{The omission of either the edge or its edge weight constitutes link missingness.} Figure \ref{fig:quality_hist} shows the histogram of the journal impact factors for published economic papers. The histogram shows clear left truncation at zero, with a substantial probability mass accumulated in the left tail, raising the question of whether truncation is the underlying mechanism. In Section \ref{sec:application}, I apply the J-test from Section \ref{sec:test} to the publication data, and provide further evidence for missing links. The presence of link missingness poses an econometric challenge for studying the \textit{collaboration} network, as only its sub-graph—the \textit{publication} network—is observed.

Estimation and inference on partially observed networks can result in misspecification error and bias if conventional estimators, which assume full network observability, are used. There are three main econometric challenges. First, the truncation mechanism introduces selection bias to the data, creating an \textit{endogenous} network $\mathcal{G}$. This means that one can no longer use standard linear regression to estimate the fixed effects and the scaling factor, as in \citet{abowd_high_1999} and \citet{bonhomme_teams_2021}, the latter implicitly assuming fully observed team networks. Second, workers and teams typically have substantial heterogeneity (\citet{bonhomme_teams_2021}) that is typically unobserved. Third, since many real-life networks are sparse (\citet{graham_econometric_2017}), the dimension of individual fixed effects grows at the same rate as the number of observations (links), a challenge further exacerbated by the link truncation. The main contribution of this paper is to provide a Generalized Method of Moment (GMM) estimator and a J-test for partially observed team networks. This paper also contributes to the empirical literature on teamwork productivity by revealing and correcting the systematic underestimation of collaboration premiums that arises from relying solely on truncated publication data. Measuring the premium is crucial, because forming teams of the optimal size can enhance productivity, \textit{ceteris paribus}. 

\subsection{Literature Review}
This paper relates to the rapidly growing literature on networks. It is often challenging to accurately collect data on social and economic networks. On the other hand, various data constraints on networks can introduce econometric complexities. When networks are constructed by a partial sample of nodes, \citet{chandrasekhar_econometrics_2011} recommend analytical correction or graphical reconstruction to alleviate estimation bias. Furthermore, \citet{chandrasekhar_non-robustness_2024} show that the estimation of the diffusion effect may not be robust even with a vanishing share of mis-measured local links. \citet{lewbel_ignoring_2024} show that the two-stage estimation and inference of the linear social effects model (\citet{manski_identification_1993}) remain valid if the measurement errors in the adjacency matrix grow sufficiently slow compared to the sample size. \citet{boucher_estimating_2023} probabilistically reconstruct partially observed networks to correct the downward bias in the estimated peer effects. For network survey data capping the maximum number of reported links, \citet{griffith_name_2022} proposes both correction and bounding methods. 

This paper also relates to the literature on limited dependent variable. Following the pioneering work of \citet{tobin_estimation_1958} on censored regression (also referred to as Tobit model), \citet{amemiya_regression_1973} establishes the consistency of Tobin's maximum likelihood estimator and provides a consistent initial estimator. Generalizing the regression bias under the limited dependent variable as a special omitted variable bias, \citet{heckman_sample_1979} constructs a consistent two-stage estimator. Moving from standard linear regression to quantile regression, the censored least absolute deviation (LAD) estimator exploits symmetry to build orthogonality conditions that are also free of parametric assumptions on the i.i.d. errors (see \citet{powell_censored_1986} and \citet{powell_symmetrically_1986}). \citet{honore_trimmed_1992} incorporates fixed effects into the censored and truncated panel models, using sample trimming to restore symmetry and build orthogonality conditions. \citet{hu_estimation_2002} focuses on a dynamic censored model in which a lagged dependent variable is introduced to the censored panel model.

The remainder of the paper is organized as follows. Section \ref{sec:model} explains the econometric model, introducing a naive estimator for fully observed networks, a GMM estimator for partially observed networks, and a test for missing links. Section \ref{sec:simulation} shows the Monte Carlo results. Section \ref{sec:application} applies the model to academic publication data. Section \ref{sec:conclusion} concludes.

\section{The Model}\label{sec:model}

\subsection{General Network Structure}
This section begins by outlining the structure of latent and observed team networks and then introduces the data-generating process (DGP). Team networks are modeled as hypergraphs, allowing links to connect any number of nodes. Self-loops, which are links that connect a node to itself, are also permitted. The superscript star $*$ is used to indicate latent variables. Denote $\mathcal{E}^*$ as the latent set of links and $\mathcal{V}^*$ as the latent set of nodes. In the latent graph $\mathcal{G^*}=(\mathcal{V}^*, \mathcal{E}^*)$, each link $\ell \in \mathcal{E}^*$ is associated with a latent outcome variable $Y_{\ell}^*$. After link truncation, the observed edge set becomes $\mathcal{E} \subseteq \mathcal{E}^*$, where each link $\ell \in \mathcal{E}$ is now associated with an observed outcome $Y_{\ell}$. Furthermore, a latent node is omitted if all of its latent links are truncated. Thus, the observed graph becomes $\mathcal{G} = (\mathcal{V}, \mathcal{E})$ where the observed node set becomes $\mathcal{V} \subseteq \mathcal{V}^*$. Notably, the model of partially observed networks \textit{encompasses} fully observed networks as a special case, where the latter has no missing links.

Because some workers may not participate in both solo and team production, let $L_i^*$ denote the latent set of self-loops, representing all solo projects undertaken by node $i$. Similarly, let $L_{ij}^*$ denote the latent set of links connecting nodes $i$ and $j$, corresponding to their joint team projects. After link truncation, the observed sets are given by $L_i \subseteq L_i^*$ for solo projects and $L_{ij} \subseteq L_{ij}^*$ for team projects.

\subsection{Team Production Model}

Consider a latent team network $\mathcal{G}^*$ where each project (link) $\ell$ is associated with a latent outcome $Y_{\ell}^* \in \RR$. However, links with $Y_\ell^* < 0$ are truncated from the graph and therefore remain unobserved. The econometrician observes $Y_{\ell}^*$ only if $Y^*_{\ell} \ge 0$, in which case the observed outcome is denoted by $Y_\ell$.\footnote{The choice of the cut-off point at zero is without loss of generality. This paper focuses on one-sided left truncation, and the right truncation case is similar.} In particular, I highlight the distinction between \textit{teams}, which consist of individual workers, and \textit{projects}, which are produced by teams. Importantly, one team may produce multiple projects, and one worker may be a member of multiple teams. 
Note that when a link is truncated, both its outcome $Y_\ell^*$ and team composition go unobserved—unlike censoring, which preserves the team composition.

Let $\alpha_i \in \mathbb{R}$ denote worker $i$'s unobserved type or individual fixed effect, and let $\alpha$ be the vector of all individual fixed effects. The index $\ell_i$ specifies which worker’s fixed effect $\alpha_i$ is associated with project $\ell$. Denote the team size of project $\ell$ as $s_\ell \in \NN^+$. The key parameter $\lambda \in (0, \infty)$, referred to as the scaling factor in \citet{bonhomme_teams_2021}, aggregates individual fixed effects $\alpha_{\ell_i}$ into a team-specific fixed effect $a_{\ell}$. Without loss of generality, the baseline model considers teams of at most two workers and a minimum team size of one; see Section \ref{sec:extensions} for extension to larger teams with $s_\ell > 2.$ The error term $U_\ell$ is scaled by an unknown standard deviation $\sigma \in (0, \infty)$. The data-generating process is:
\begin{align} \label{model:baseline}
    &a_\ell = 
    \begin{cases}
        \alpha_{\ell_i}, & \text{ if } s_\ell = 1, \\
        \lambda (\alpha_{\ell_i} + \alpha_{\ell_j}), & \text{ if }  s_\ell = 2, 
    \end{cases}
    \nonumber \\
    &Y^*_{\ell} = a_{\ell} + \sigma U_{\ell}, 
    \nonumber \\
    & Y_{\ell} =
    \begin{cases}
        Y^*_{\ell}, & \text{if } Y^*_{\ell} \geq 0, \\
        \textnormal{Unobserved}, & \text{if } Y^*_{\ell} < 0.
    \end{cases}
\end{align}
For single-worker teams, the team fixed effect reduces to the worker-specific fixed effect, which is simply $\alpha_{\ell_i}$. Here, the scaling factor for one-worker teams is normalized to unity. When the team has two workers, its fixed effect becomes $\lambda (\alpha_{\ell_i} + \alpha_{\ell_j})$. This additive specification implies that team production is symmetric between contributors—the order of workers does not matter. 

Furthermore, the scaling factor $\lambda$ plays a similar role to the parameter $\beta_n$ in the generalized Hölder means specification of \citet{ahmadpoor_decoding_2019}, where $\beta_n$ captures the impact benefits of $n$-person teams. It governs the sign and magnitude of collaboration premium because incorporating $\lambda$ flexibly enables collaboration to have productivity gain (or loss) relative to solo production. In contrast, the arithmetic mean assumes no collaboration premium, as it simply scales the sum of worker fixed effects by the inverse of the team size. For instance, teams of two and three people use fixed weights of $\frac{1}{2}$ and $\frac{1}{3}$, respectively, both implying zero productivity gain or loss from collaboration. This structure also keeps the fixed effects dimension manageable by expressing team fixed effects as linear combinations of individual worker fixed effects. I now turn to the main assumption on the error term.

\begin{assumption}[Conditionally IID Normal Shocks]
    \label{assumption:iid}
    Conditional on the vector of individual fixed effects $\alpha$ and the latent graph $\mathcal{G}^*$, the project-specific shocks $\{U_\ell\}$ are independently and identically distributed as standard normal:
    \[
    U_\ell \simiid N(0, 1) \,|\, \alpha, \, \mathcal{G}^*.
    \]
\end{assumption}

Section \ref{sec:extensions} extends the model by allowing the variance parameter $\sigma_\ell$ to differ across team sizes $s_\ell$. It also discusses how to incorporate time-varying covariates and how to generalize the framework to accommodate larger teams. However, the current model does not incorporate a team formation model or allow for peer effects or spillover among workers. Furthermore, I assume no serial dependence between repeated teamwork. This assumption is strong but plausible in innovative settings where workers study novel ideas, ensuring project-level shocks remain uncorrelated.

\begin{remark}
    Throughout this paper, all stochastic specifications are implicitly conditioned on the latent graph $\mathcal{G}^*$ and fixed effects $\alpha$. This allows arbitrary dependence between team formation and individual heterogeneity.
\end{remark}

\subsection{Fully Observed Network}\label{sec:estimator_naive}
Before addressing the general framework of partially observed team networks, I first discuss a special case where the latent team network $\mathcal{G}^*$ is fully observed. In other words, $Y_\ell^*$ is observed for all $\ell$, and $\mathcal{G} = \mathcal{G}^*$ holds. While the fully observed network is not the primary focus of this paper, it helps illustrate the endogeneity problem and the limitations of traditional methods. I will return to the model of partially observed networks in Section \ref{sec:GMM}. Under full network observability, I consider a team production model studied in \citet{bonhomme_teams_2021}:
\begin{align} \label{model:fully_observed}
    &a_\ell = 
    \begin{cases}
        \alpha_{\ell_i}, & \text{ if } s_\ell = 1, \\
        \lambda (\alpha_{\ell_i} + \alpha_{\ell_j}), & \text{ if }  s_\ell = 2, 
    \end{cases}
    \nonumber \\
    &Y_{\ell} = a_{\ell} + \sigma U_{\ell}.
\end{align}
Since every link is observed regardless of its associated outcome, $Y_\ell = Y_{\ell}^*$ for all $\ell$, which eliminates the need to use latent variables. One can thus relax the parametric assumption in Assumption (\ref{assumption:iid}), and instead impose a weaker assumption on the error term.
\begin{assumption}[Mean-zero shock]
    \label{assumption:mean_zero}
    The shock $U_\ell$ is mean-zero conditional on $\alpha$ and $\mathcal{G}^*$, with
    $\EE[U_\ell \mid \alpha, \, \mathcal{G}^*] = 0$.
\end{assumption}

To derive a consistent estimator of the scaling factor, I compare the expected outcome of a team project to the combined expected outcomes of the solo projects, weighted by the unknown parameter $\lambda$. Specifically, I subtract the unconditional expectation of $\lambda(Y_{\ell_i}^* + Y_{\ell_j}^*)$ from that of $Y_{\ell_{ij}}^*$, which yields the following moment condition: 
\begin{equation}\label{moment:linear_latent}
    \EE[Y_{\ell_{ij}}^* - \lambda (Y_{\ell_i}^* + Y_{\ell_j}^*) \,;\, \ell_i \in L_i^*, \ell_j \in L_j^*, \ell_{ij} \in L_{ij}^*] = 0.
\end{equation}
This condition implies that identification requires at least one solo project for each worker and at least one team project linking workers $i$ and $j$. Because $\lambda$ enters linearly, it admits a closed-form estimator that takes the form of a simple ratio.
\begin{lemma} \label{eq:naive}
    Suppose the latent network $\mathcal{G}^*$ is fully observed and Assumption (\ref{assumption:mean_zero}) holds. Define the average solo-project outcome for worker $i$ and the average team-project outcome between workers $i$ and $j$ as
    \[
        \overline{Y_i}^* = \frac{1}{|L_{i}^*|} \sum_{\ell \in L_i^*} Y_{\ell}^*, \quad
        \overline{Y_{ij}}^* = \frac{1}{|L_{ij}^*|} \sum_{\ell \in L_{ij}^*} Y_{\ell}^*.
    \]
    Then, the estimator
    \begin{equation}
        \hat\lambda_{\text{naive}}(\mathcal{G}^*) = \frac{\sum\limits_{i,j \in \mathcal{V}^*,\, i<j} \mathbbm{1}\{|L_{i}^*| > 0, |L_{j}^*| > 0, |L_{ij}^*| > 0\} \overline{Y_{ij}}^*}{\sum\limits_{i,j \in \mathcal{V}^*,\, i<j} \mathbbm{1}\{|L_{i}^*| > 0, |L_{j}^*| > 0, |L_{ij}^*| > 0\} (\overline{Y_i}^*+ \overline{Y_j}^*)}
    \end{equation} 
    is a consistent estimator of $\lambda$.
\end{lemma}

This estimator compares team output to solo output, implicitly controlling for unobserved worker types. A related idea appears in \citet{anderson_collaborative_2022}, who subtract each researcher’s average solo productivity from team outcomes to isolate team-specific factors—such as team size, information spillovers, and time allocation—using a regression-tree model. Here I focus solely on the team-size effect and assume no network spillovers. A key benefit of the naive estimator is that it avoids parametric assumptions on the error term beyond a zero conditional mean.

Without link truncation, that is, $\mathcal{G} = \mathcal{G}^*$, this naive estimator is correctly specified. However, with link truncation, projects with negative outcomes ($Y_\ell^* < 0$) become unobserved so that $\EE[U_\ell \mid \alpha,\, \mathcal{G}] \ne 0$. Intuitively, the truncation mechanism favors projects with larger $U_\ell$. For instance, conditional on the same worker types, ``luckier'' projects are more likely to reach the publication stage. As a result, the moment equality in (\ref{moment:linear_latent}) breaks down on the truncated network. An immediate consequence is that $\hat\lambda_{\text{naive}}(\mathcal{G})$ is inconsistent, which in turn biases estimates of $\sigma$ and, when relevant, the latent types $\alpha_i$. Thus, an alternative estimator is required. Nevertheless, the naive estimator remains useful: it will serve as the basis for constructing a test for missing links in Section \ref{sec:test}.

\subsection{Partially Observed Network}\label{sec:GMM}
The key innovation in the team production model—relative to \citet{bonhomme_teams_2021}—is the incorporation of link truncation, which introduces endogeneity when networks are only partially observed. The goal is to construct a conditional moment function $m(\lambda, \sigma, Y_{\ell_i}, Y_{\ell_j}, Y_{\ell_{ij}})$ such that
\begin{align*}
    \mathbb{E}\left[
    m(\lambda, \sigma, Y_{\ell_i}, Y_{\ell_j}, Y_{\ell_{ij}}) \,;\, \ell_i \in L_i^*, \ell_j \in L_j^*, \ell_{ij} \in L_{ij}^*
    \mid 
    Y_{\ell_i}^* \ge 0, Y_{\ell_j}^* \ge 0, Y_{\ell_{ij}}^* \ge 0
    \right] = 0 
\end{align*}
for all possible values of individual fixed effects. Conditioning on non-negative project outcomes, $Y_{\ell_i}^* \ge 0, \, Y_{\ell_j}^* \ge 0, \, Y_{\ell_{ij}}^* \ge 0$, ensures that this moment condition remains valid even when some links or nodes are systematically omitted from the observed network.

One strategy is to impose a parametric assumption on the error term to structure the missingness. \citet{tobin_estimation_1958} pioneered the use of normal errors in limited-dependent-variable models. Building on this, \citet{amemiya_regression_1973} used the first four moments to construct a consistent initial estimator for Tobit models, while \citet{honore_iv_1998} exploited the relationship between the first two moments of the truncated normal to difference out individual fixed effects and develop an IV estimator for censored panel data. \citet{horrace_moments_2015} studied properties of the first four truncated-normal moments, and \citet{orjebin_recursive_2014} derived a recursive formula for these moments. I do not take credit for the lemma below.

\begin{lemma} \label{lemma:TruncatedNormal}
    Suppose $\widetilde{Y} \sim N(\tilde{\alpha},\, \tilde{\sigma}^2)$ where $\tilde{\alpha} \in \RR$ and $\tilde{\sigma} \ge 0$. Then, for $k \in \NN^+$,
    \[
    \EE[\widetilde{Y}^{k+1} - \tilde{\alpha} \widetilde{Y}^{k} - k \tilde{\sigma}^2 \,\widetilde{Y}^{k-1} | \widetilde{Y} \ge 0] = 0.
    \]
\end{lemma}

The key insight is to leverage the normality assumption to express the higher-order conditional moments of $Y_{\ell}^*$ as a linear combination of the lower-order conditional moments and the model parameters, eliminating the dependency on missing data. A remaining challenge is that the dimension of individual fixed effects $\alpha$ grows at the same rate as the number of observations. To address this, I need to take care of the incidental parameter problem, as discussed in the seminal paper by \citet{neyman_consistent_1948}, to ensure estimation consistency. The decomposition in Lemma \ref{lemma:TruncatedNormal} is particularly useful because the two model parameters $(\tilde{\alpha},\, \tilde{\sigma})$ enter the moment equation linearly in separate terms. This enables us to difference out both node-level and link-level fixed effects, allowing us to build moment conditions free of fixed effects.

To illustrate the differencing approach, set $k=1$ in Lemma~(\ref{lemma:TruncatedNormal}) and apply it to the triplet $(Y_{\ell_i},Y_{\ell_j},Y_{\ell_{ij}})$. This forms a system of two individual-related moments and one team-related moment. Although the team-related moment $Y_{\ell_{ij}}$ incorporates the parameter of interest $\lambda$, it is also entangled with the nuisance terms $\lambda(\alpha_{\ell_i} + \alpha_{\ell_j})$. Scaling $Y_{\ell_i}$ and $Y_{\ell_j}$ by $\lambda$ recovers $(\lambda\alpha_i, \lambda\alpha_j)$ and yields transformed variables $(\lambda Y_{\ell_i}, \lambda Y_{\ell_j})$ that are free of fixed effects. Because projects are mutually independent, I can further scale these moments using suitable products of $Y_{\ell_i}$, $Y_{\ell_j}$, and $Y_{\ell_{ij}}$. By jointly conditioning on $Y_{\ell_i}^*, Y_{\ell_j}^*, Y_{\ell_{ij}}^* \ge 0$, the team-related moment can then be differenced against the sum of the individual-related moments, yielding a conditional equality involving only parameters and observed data.
\begin{proposition} \label{prop:moments}
    Define the $k^{\text{th}}$ moment condition $m_k$ as
    $$
    m_k(\lambda, \sigma, Y_{\ell_i}, Y_{\ell_j}, Y_{\ell_{ij}}) := 
    Y_{\ell_i}^k Y_{\ell_j}^k Y_{\ell_{ij}}^k (Y_{\ell_{ij}} - \lambda(Y_{\ell_i} + Y_{\ell_j})) + k\sigma^2  Y_{\ell_i}^{k-1} Y_{\ell_j}^{k-1} Y_{\ell_{ij}}^{k-1} (\lambda(Y_{\ell_i} + Y_{\ell_j})Y_{\ell_{ij}}  -  Y_{\ell_{i}} Y_{\ell_{j}}).
    $$
    Then, for $k \in \NN^+$,
    $$
    \EE\left[
    m_k(\lambda, \sigma, Y_{\ell_i}, Y_{\ell_j}, Y_{\ell_{ij}}) 
    \mid
    Y_{\ell_i}^* \ge 0, Y_{\ell_j}^* \ge 0, Y_{\ell_{ij}}^* \ge 0
    \right] = 0.
    $$
\end{proposition}

Because these moments are free of the fixed effects, they are well suited to team networks in which many nodes have few links. In principle, an infinite sequence of conditional moments is available; in practice, higher-order moments are less robust because they rely more heavily on the parametric structure. Since the baseline model has only two unknown parameters $(\lambda,\sigma)$, we therefore construct conditional moments using $k \in \{1, 2\}$.
\begin{equation}
    m(\lambda, \sigma, Y_{\ell_i}, Y_{\ell_j}, Y_{\ell_{ij}})' :=
    \begin{pmatrix}
        m_1(\lambda, \sigma, Y_{\ell_i}, Y_{\ell_j}, Y_{\ell_{ij}}) &
        m_2(\lambda, \sigma, Y_{\ell_i}, Y_{\ell_j}, Y_{\ell_{ij}})
    \end{pmatrix}.
\end{equation}

Since triplets that share a project are correlated by construction, enumerating all admissible triplets induces cross-triplet dependence. To avoid this, Theorem~\ref{theorem:gmm}—assuming its conditions hold—applies GMM to a subsample of uncorrelated triplets constructed via Algorithm~\ref{alg:triplet_construction}, where no project is reused to ensure triplet independence. Including correlated triplets is feasible, but requires (i) adjusting the asymptotic covariance to account for cross-triplet dependence and (ii) imposing additional regularity conditions on the graph (to rule out extreme dependence) to ensure convergence.

\begin{theorem} \label{theorem:gmm}
    Let $B$ be the set of uncorrelated triplets $\{(Y_{\ell_i}, Y_{\ell_j}, Y_{\ell_{ij}})\,;\, \ell_i \in L_i, \ell_j \in L_j, \ell_{ij} \in L_{ij}\}$. Let $\EE^\dagger$ and $\Var^\dagger$ denote conditional expectation and conditional variance given $Y_{\ell_i}^*, Y_{\ell_j}^*, Y_{\ell_{ij}}^* \ge 0$. Define
    \[
    \hat g(\lambda, \sigma)' = \frac{1}{|B|} \sum\limits_{\ell_i, \ell_j, \ell_{ij}:(Y_{\ell_i}, Y_{\ell_j}, Y_{\ell_{ij}}) \in B}
    \begin{pmatrix}
        m_1(\lambda, \sigma, Y_{\ell_i}, Y_{\ell_j}, Y_{\ell_{ij}}) &
        m_2(\lambda, \sigma, Y_{\ell_i}, Y_{\ell_j}, Y_{\ell_{ij}})
    \end{pmatrix}.
    \]
    Suppose
    \begin{enumerate}[(i)]
        \item Assumption (\ref{assumption:iid}) holds.
        \item The limiting gradient and variance matrices exist. Specifically,
        \begin{align*}
            &G = \plim\limits_{|B| \to \infty} \frac{1}{|B|} \sum\limits_{\ell_i, \ell_j, \ell_{ij}:(Y_{\ell_i}, Y_{\ell_j}, Y_{\ell_{ij}}) \in B} 
            \EE^\dagger
            \begin{bmatrix}
            \frac{\partial}{\partial \lambda} m(\lambda, \sigma, Y_{\ell_i}, Y_{\ell_j}, Y_{\ell_{ij}}), &
            \frac{\partial}{\partial \sigma} m(\lambda, \sigma, Y_{\ell_i}, Y_{\ell_j}, Y_{\ell_{ij}})
            \end{bmatrix},
            \\
            &V = \plim\limits_{|B| \to \infty} \frac{1}{|B|} \sum\limits_{\ell_i, \ell_j, \ell_{ij}:(Y_{\ell_i}, Y_{\ell_j}, Y_{\ell_{ij}}) \in B}
            \Var^\dagger(m(\lambda, \sigma, Y_{\ell_i}, Y_{\ell_j}, Y_{\ell_{ij}})).
            \end{align*}
        \item Global identification of $(\lambda,\, \sigma)$.
    \end{enumerate}
    Then, the baseline model (\ref{model:baseline}) has a consistent GMM estimator, that is,
        \begin{equation*}
        \begin{pmatrix}
            \hat\lambda \\ \hat\sigma
        \end{pmatrix} 
        = 
        \argmin_{\lambda \in \RR, \sigma \in \RR^+} 
        \hat g(\lambda, \sigma)' \hat g(\lambda, \sigma).
        \end{equation*}
        Furthermore, as $|B| \to \infty$, the asymptotic distribution is
        \begin{equation*}
            \sqrt{|B|}
                \left[
                \begin{pmatrix}
                \hat\lambda \\ \hat\sigma
                \end{pmatrix} 
                -
                \begin{pmatrix}
                \lambda \\ \sigma
                \end{pmatrix}
                \right]
                \Longrightarrow
                N\left(0, (G' V^{-1} G)^{-1}\right).     
        \end{equation*}
\end{theorem}

\begin{remark}[Rank condition for global identification] \label{remark:rank}
    To verify identification, it is sufficient to check the rank of the following linear system.\footnote{Note that there may potentially exist a weaker identification condition.} Consider the parameter vector $(\lambda, \sigma^2, \rho)$ with $\rho := \lambda \sigma^2$. Applying Proposition \ref{prop:moments} for $k \in \{1,2,3\}$ yields three moment equations, which can be written as
    \[
    \EE^\dagger
    \left(
    \begin{bmatrix}
    - (Y_{\ell_i} + Y_{\ell_j})Y_{\ell_{i}} Y_{\ell_{j}}Y_{\ell_{ij}} & -Y_{\ell_{i}}Y_{\ell_{j}} & (Y_{\ell_i} + Y_{\ell_j})Y_{\ell_{ij}}\\
    - (Y_{\ell_i} + Y_{\ell_j})Y_{\ell_{i}}^2 Y_{\ell_{j}}^2Y_{\ell_{ij}}^2 &  -2Y_{\ell_{i}}^2Y_{\ell_{j}}^2Y_{\ell_{ij}} & 2(Y_{\ell_i} + Y_{\ell_j})Y_{\ell_{i}} Y_{\ell_{j}} Y_{\ell_{ij}}^2 \\
    - (Y_{\ell_i} + Y_{\ell_j})Y_{\ell_{i}}^3 Y_{\ell_{j}}^3Y_{\ell_{ij}}^3 
    & -3Y_{\ell_{i}}^3Y_{\ell_{j}}^3Y_{\ell_{ij}}^2 & 3(Y_{\ell_i} + Y_{\ell_j})Y_{\ell_{i}}^2 Y_{\ell_{j}}^2Y_{\ell_{ij}}^3\\
    \end{bmatrix}
    \begin{bmatrix}
    \lambda \\
    \sigma^2 \\
    \rho
    \end{bmatrix}
    +
    \begin{bmatrix}
    Y_{\ell_{i}} Y_{\ell_{j}} Y_{\ell_{ij}}^2 \\
    Y_{\ell_{i}}^2 Y_{\ell_{j}}^2 Y_{\ell_{ij}}^3 \\
    Y_{\ell_{i}}^3 Y_{\ell_{j}}^3 Y_{\ell_{ij}}^4
    \end{bmatrix}
    \right)
    =0.
    \] 
    Therefore, a full rank of the leftmost matrix ensures identification of $(\lambda, \sigma)$.
\end{remark}

If desired, one can also recover unbiased estimates of the individual fixed effects $\alpha_i$. There are two main sources of bias in estimating the fixed effects: bias in the scaling factor and truncation bias. To illustrate the latter, consider a worker with a negative fixed effect. If links are truncated at zero, then under a standard regression, their estimated type will always appear non-negative — introducing systematic upward bias. After applying Theorem \ref{theorem:gmm} to obtain consistent estimates $(\hat\lambda,\hat\sigma)$, Lemma \ref{lemma:TruncatedNormal} implies moments for estimating an identified subset of $\alpha$. For each worker $i$ with solo project $\ell_i\in L_i$, $\EE\left[Y_{\ell_i}^{2}-\alpha_i Y_{\ell_i}-\hat{\sigma}^{2}\, |\,Y_{\ell_i}^{*}\ge 0\right]=0$, contributing $\sum\limits_{i\in\mathcal V}\lvert L_i\rvert$ moments; and for each unordered team $\{i,j\}$ with team project $\ell_{ij}\in L_{ij}$, $\EE[Y_{\ell_{ij}}^{2}-\hat{\lambda}(\alpha_i+\alpha_j)Y_{\ell_{ij}}-\hat{\sigma}^{2}\,|\,Y_{\ell_{ij}}^{*}\ge 0]=0$, yielding an additional $\sum\limits_{i,j \in \mathcal{V}, i<j} |L_{ij}|$ moments.

\begin{remark}[Noise in the estimated fixed effects]
    To meaningfully interpret the fixed-effect estimates, which might be estimated with significant noises, it is important to verify the network is sufficiently dense and connected. A dense team network implies that every worker has, on average, a relatively large number of observations. Moreover, greater network connectivity enhances estimation precision, as formalized by \citet{jochmans_fixed-effect_2019} in the context of individual fixed-effect estimation in a two-way regression model.
\end{remark}

\subsection{Extensions} \label{sec:extensions}

\subsubsection{Heteroscedasticity}
Heterogeneous variances can be accommodated in the baseline model. While the setup in \eqref{model:baseline} assumes conditionally i.i.d. shocks across projects $\ell$, one may allow the variance parameter $\sigma_{s_\ell}$ to vary across team sizes $s_\ell \in \NN^+$. For instance, team output may exhibit greater (or lower) volatility than solo output. Let $(\sigma_1, \sigma_2)$ denote the variances of the one-worker and two-worker production, respectively. The extended model becomes:
\begin{align} \label{model:heterog_var}
    &a_\ell = 
    \begin{cases}
        \alpha_{\ell_i}, & \text{ if } s_\ell = 1, \\
        \lambda (\alpha_{\ell_i} + \alpha_{\ell_j}), & \text{ if }  s_\ell = 2, 
    \end{cases}
    \nonumber \\
    &Y^*_{\ell} = a_{\ell} + \sigma_{s_\ell} U_{\ell}, 
    \nonumber \\
    & Y_{\ell} =
    \begin{cases}
        Y^*_{\ell}, & \textnormal{if } Y^*_{\ell} \geq 0, \\
        \textnormal{Unobserved}, & \textnormal{if } Y^*_{\ell} < 0.
    \end{cases}
\end{align}
To estimate $(\lambda, \sigma_1, \sigma_2)$, one can simply apply Proposition \ref{prop:moments} with $k=1, 2, 3$. Now, with three moment restrictions, the parameters are just-identified.

\subsubsection{Covariates}

In some cases, it may be important to incorporate covariates into the team production function. Time-invariant covariates are absorbed by the individual fixed effects $\alpha_i$ and are not identified. In contrast, time-varying covariates $X_{\ell, i, t}$ are not absorbed. For instance, suppose we estimate $\beta \in \RR$, the marginal effect of one additional year of work experience on paper quality. Define $X_{\ell, i, t}$ as the academic work experience of researcher $i$ at the time of writing paper $\ell$, measured as the number of years since graduation. The model becomes:

\begin{align} \label{model:covariates}
    &a_\ell = 
    \begin{cases}
        \alpha_{\ell_i}, & \text{ if } s_\ell = 1, \\
        \lambda (\alpha_{\ell_i} + \alpha_{\ell_j}), & \text{ if }  s_\ell = 2, 
    \end{cases}
    \nonumber \\
    &Y^*_{\ell, t} =    
    \begin{cases}
        a_{\ell} + \beta X_{\ell, i, t} + U_{\ell, t}, & \text{ if } s_\ell = 1, \\
        a_{\ell} + \beta \lambda \left(X_{\ell, i, t} + X_{\ell, j, t}\right) + U_{\ell, t}, & \text{ if }  s_\ell = 2,
    \end{cases}
    \nonumber \\
    & Y_{\ell, t} =
    \begin{cases}
        Y^*_{\ell, t}, & \textnormal{if } Y^*_{\ell, t} \geq 0, \\
        \textnormal{Unobserved}, & \textnormal{if } Y^*_{\ell, t} < 0.
    \end{cases}
\end{align}

The parameter vector is now $(\lambda,\, \sigma,\, \beta)$, and each project $\ell$ carries both a worker index $i$ and a time index $t$. Note that while the covariates enter the production function additively, they are multiplied by $\beta$ and $\lambda$. Proposition \ref{prop:moments} still applies, because the covariates components simply shift the mean. To identify $\beta$, sufficient temporal variation is needed so that the change in $i$'s characteristics at the time of producing $i$'s solo project and the team project must \textit{not} exactly offset the change in $j$'s characteristics at the time of producing $j$'s solo project and the team project. Otherwise, all the time-varying components cancel out, along with $\beta$. I therefore apply Proposition \ref{prop:moments} with $k = 1, 2, 3$ to derive a similar GMM estimator. Additionally, project-level covariates (e.g., grant amounts) are straightforward to include.

\subsubsection{Three or More Workers}
Although the baseline DGP considers at most two workers per team, the model extends naturally to teams of arbitrary size. To accommodate this extension, we index the scaling factor by team size $s \in \NN^+$, denoting it by $\{\lambda_s\}$ and normalizing the scaling factor for single-worker teams $\lambda_1$ to unity, consistent with the baseline model. For example, when the team-size is three, the team-specific fixed effect $a_\ell$ becomes a weighted average of types of workers $i$, $j$ and $k$. Specifically, $a_\ell = \lambda_3 (\alpha_{\ell_i} + \alpha_{\ell_j} + \alpha_{\ell_k})$ where $\lambda_3 \in \RR$ is the scaling factor associated with three-person teams. Extending \eqref{model:baseline}, the model becomes:
\begin{align} \label{model:larger_teams} 
    &a_\ell = 
    \begin{cases}
        \alpha_{\ell_i}, & \text{ if } s_\ell = 1, \\
        \lambda_2 (\alpha_{\ell_i} + \alpha_{\ell_j}), & \text{ if }  s_\ell = 2, \\
        \lambda_3 (\alpha_{\ell_i} + \alpha_{\ell_j} + \alpha_{\ell_k}), & \text{ if }  s_\ell = 3,
    \end{cases}
    \nonumber \\
    &Y^*_{\ell} = a_{\ell} + \sigma U_{\ell}, 
    \nonumber \\
    & Y_{\ell} =
    \begin{cases}
        Y^*_{\ell}, & \text{if } Y^*_{\ell} \geq 0, \\
        \textnormal{Unobserved}, & \text{if } Y^*_{\ell} < 0.
    \end{cases}
\end{align}
The GMM estimator now incorporates the four-dimensional tuple $(Y_{\ell_i}, Y_{\ell_j}, Y_{\ell_k}, Y_{\ell_{ijk}})$, and Proposition \ref{prop:moments} is applied again to construct a new moment involving $\lambda_3$ but free of fixed effects.

\subsection{Test for Missing Links}\label{sec:test}

Before selecting an estimator, one naturally asks whether the network data at hand is only partially observed. Network observability can be evaluated using applied and theoretical knowledge, or through statistical tests.\footnote{It may be necessary to employ robust post-selection adjustments to mitigate pre-testing bias.} However, due to the nature of missing data, directly testing for missing links is generally infeasible. Even if the empirical distribution appears truncated—for instance, in Figure \ref{fig:quality_hist}—the observed accumulation of probability mass, however, could alternatively arise from an idiosyncratic data-generating process with bounded support. This section introduces a test for link truncation that avoids any distributional assumption on the error term. Consider the null hypothesis $\mathcal{H}_0: \mathcal{G} = \mathcal{G}^*$ against $\mathcal{H}_1: \mathcal{G} \subsetneq \mathcal{G}^*$.

In the GMM framework, the J-test provided in \citet{hansen_large_1982} provides a simple way to test the validity of moment functions. The goal is to construct additional over-identifying moments—beyond (\ref{moment:linear_latent}) implied by the DGP—that hold under $\mathcal{H}_0$ but fail under $\mathcal{H}_1$. To achieve this, I exploit the transition from the exogenous network $\mathcal{G}^*$, which holds under $\mathcal{H}_0$, to the endogenous network $\mathcal{G}$, which holds under $\mathcal{H}_1$. To see this, recall that Assumption (\ref{assumption:iid}) implies $U \notindependent \mathcal{G} \,|\, \alpha$. However, when a link $\ell$ with $Y_{\ell}^* <0$ is omitted, the shock $U_{\ell}$ becomes positively correlated with the indicator of observing $\ell$, which is $\mathbf{1}{\{Y_\ell^* \ge 0\}}$.

Leveraging the network exogeneity under the null, additional moments can be generated by interacting (\ref{moment:linear_latent}) with arbitrary network statistics (e.g., node degrees) computed at the team level to match each triplet’s granularity. Rejection of these moments in the J-test below implies rejection of the null, allowing truncation to be tested even without observing the omitted links.

\begin{lemma}[J-test for missing links]\label{J_test}
     Let the null hypothesis $\mathcal{H}_0$ state that no links are missing. Let $C \subseteq B$ denote the subset of uncorrelated triplets such that each node pair $\{i, j\}$ with $i, j \in \mathcal{V}$ appears exactly once. For each $\{i, j\}$, compute network statistics $f_{ij}^k(\mathcal{G})$ for $k = 1, \ldots, K$. Define
     \begin{align*}
     &m_{0}(\lambda, Y_{\ell_i}, Y_{\ell_j}, Y_{\ell_{ij}}) = Y_{\ell_{ij}} - \lambda (Y_{\ell_i} + Y_{\ell_j}),
     \\
     &g_0(\lambda) = \frac{1}{|C|} \sum\limits_{\ell_i, \ell_j, \ell_{ij}:(Y_{\ell_i}, Y_{\ell_j}, Y_{\ell_{ij}}) \in C} m_0(\lambda, Y_{\ell_i}, Y_{\ell_j}, Y_{\ell_{ij}}),
     \\
     &g_k(\lambda) = \frac{1}{|C|} \sum\limits_{\ell_i, \ell_j, \ell_{ij}:(Y_{\ell_i}, Y_{\ell_j}, Y_{\ell_{ij}}) \in C} f_{ij}^k(\mathcal{G}) \cdot m_{0}(\lambda, Y_{\ell_i}, Y_{\ell_j}, Y_{\ell_{ij}}), \quad k = 1, \ldots, K.
     \end{align*}
     Define
     $
     g(\lambda)' := \begin{pmatrix} g_0(\lambda) & g_1(\lambda) & \cdots & g_K(\lambda) \end{pmatrix}.
     $
     Under $\mathcal{H}_0$, let $\hat S_{\mathcal{H}_0}$ and $\hat\lambda$ denote consistent estimators of the asymptotic covariances of $S$ and $\lambda$, respectively. Then, as $|C| \rightarrow \infty$, the J-statistic
     \begin{align*}
     T_{\mathcal{H}_0} = |C|\, g(\hat\lambda)' \, {\hat S_{\mathcal{H}_0}}^{-1} \, g(\hat\lambda) \sim \chi^2_K. 
     \end{align*} 
\end{lemma}
To see why $f_{ij}^k(\mathcal{G}) \cdot m_0(\lambda, Y_{\ell_i}, Y_{\ell_j}, Y_{\ell_{ij}})$ is a valid moment, note that $U \independent \mathcal{G} \,|\, \alpha$ under the null, so $f_{ij}^k(\mathcal{G}) \independent U \,|\, \alpha$. Hence, $\EE[f_{ij}^k(\mathcal{G}) \,\cdot\, m_0(\lambda, Y_{\ell_i}, Y_{\ell_j}, Y_{\ell_{ij}})] = 0$. However, one should still carefully explore and select appropriate network statistics, as some statistics are more informative than others and can thus enhance the statistical power of the J-test. Finally, one caveat applies to the interpretation of this test. Because the over-identifying restrictions rely on a correctly specified production function, rejection of the null may reflect either missing links or misspecified linear team production function.

\section{Simulation} \label{sec:simulation}
To numerically assess the GMM estimator, I conduct Monte Carlo simulations to test its performance under small-sample, sparsity and misspecified error distributions. The simulations focus on the scaling factor $\lambda$, with true parameter values set to $\lambda = 0.7$ and $\sigma = 2$. For identification, each node is required to have at least one solo project and one team project. Algorithm \ref{alg:simulation}, provided in the Appendix, is repeated 1,000 times for each simulation setting.

\subsection{Sparsity}
Many real-world networks are sparse. A network is said to be sparse when the ratio of observed links, $|\mathcal{E}|$, to all possible pairs, $|\mathcal{V}|^2$, approaches zero as the number of nodes grows.\footnote{In a multigraph, sparsity reflects both the fraction of connected node pairs and the number of parallel links within each pair.} It is therefore important to assess the performance of the GMM estimator under sparse conditions. Because the estimator relies only on triplets $(Y_{\ell_{ij}}, Y_{\ell_i}, Y_{\ell_j})$ rather than the full graph, it should, in principle, remain robust in sparse networks.

In the simulation, network sparsity is controlled by varying the average number of links per node. The leftmost panel of Table \ref{table:sparsity} reports results for a highly sparse network with one-tenth of a link per node on average. The middle panel corresponds to a moderately sparse case with roughly one link per node, while the rightmost panel presents a denser network averaging ten links per node. As expected, the variances of both the naive and GMM estimators decline with increasing density due to the larger number of observations. The naive estimator remains efficient and consistent under full observability; however, its substantial downward bias persists under missing links, even in denser networks. In contrast, the GMM estimator markedly reduces both the median bias and the median absolute error (MAE).

\begin{table}[!h]
   \centering
   \resizebox{\columnwidth}{!}{
    \begin{tabular}{llrrrrrrrrrrrr}
\toprule
 &  & \multicolumn{4}{r}{Highly Sparse (\# Links = 0.1 \# Nodes)} & \multicolumn{4}{r}{Sparse (\# Links = \# Nodes)} & \multicolumn{4}{r}{Less Sparse (\# Links = 10 \# Nodes)} \\
 &  & True Value & Bias & MAE & SE & True Value & Bias & MAE & SE & True Value & Bias & MAE & SE \\
\cmidrule(lr){3-6} \cmidrule(lr){7-10} \cmidrule(lr){11-14}
\midrule
\multirow[t]{2}{*}{Fully Observed Network} & Naive Estimator & 70.00\% & -0.10\% & 1.20\% & 1.78\% & 70.00\% & 0.01\% & 0.39\% & 0.58\% & 70.00\% & 0.00\% & 0.10\% & 0.16\% \\
 & GMM Estimator & 70.00\% & -0.25\% & 2.80\% & 4.13\% & 70.00\% & -0.25\% & 1.15\% & 1.65\% & 70.00\% & -0.07\% & 0.38\% & 0.54\% \\
\cline{1-14}
\multirow[t]{2}{*}{Partially Observed Network} & Naive Estimator & 70.00\% & -6.96\% & 6.96\% & 1.37\% & 70.00\% & -6.94\% & 6.94\% & 0.44\% & 70.00\% & -6.94\% & 6.94\% & 0.15\% \\
 & GMM Estimator & 70.00\% & -0.16\% & 4.16\% & 6.15\% & 70.00\% & 0.43\% & 3.50\% & 6.22\% & 70.00\% & 0.08\% & 1.79\% & 2.72\% \\
\cline{1-14}
\bottomrule
\end{tabular}
}
    \caption{Simulation: Sparsity}
    \label{table:sparsity}
    \floatfoot{\footnotesize{\textit{Notes: This table reports simulation results for the naive and GMM estimator when the graph is highly sparse, sparse and less sparse. The notation — $\# \text{Links} = r \, \# \text{Nodes}$ — implies a link-to-node ratio of $r$. I fix the number of nodes at $10,000$ to mimic the size of the economics publication network. At the same time, I set the total number of links to $1,000$, $10,000$ and $100,000$, to simulate highly sparse, sparse, and less sparse settings, respectively. Bias refers to the median bias, and MAE refers to the median absolute error. The standard error (SE) is computed as the interquartile range of the simulated estimates divided by $1.35$. I simulate $1,000$ times.}}}
\end{table}

\subsection{Size}

Social and economic networks typically differ in size. This section considers three settings—small, medium, and large—while maintaining a sparse structure where the number of nodes equals the number of links. As network size decreases, both estimators become more sensitive to sampling variation, reflected in larger standard errors. For fully observed networks, the naive estimator performs better in terms of bias and efficiency. In contrast, under partial observability, the GMM estimator outperforms the naive estimator by correcting the downward bias in the scaling factor. As the network grows, the GMM estimator reduces bias by an order of magnitude relative to the naive estimator, whereas the latter shows little improvement. Intuitively, because the GMM estimator relies on locally defined triplets, a larger network provides more such observations, improving precision without being hindered by sparsity.

\begin{table}[!h]
   \centering
   \resizebox{\columnwidth}{!}{
    \begin{tabular}{llrrrrrrrrrrrr}
\toprule
 &  & \multicolumn{4}{r}{Small-sized (100 nodes)} & \multicolumn{4}{r}{Medium-sized (1,000 nodes)} & \multicolumn{4}{r}{Large-sized (10,000 nodes)} \\
 &  & True Value & Bias & MAE & SE & True Value & Bias & MAE & SE & True Value & Bias & MAE & SE \\
\cmidrule(lr){3-6} \cmidrule(lr){7-10} \cmidrule(lr){11-14}
\midrule
\multirow[t]{2}{*}{Fully Observed Network} & Naive Estimator & 70.00\% & -0.08\% & 3.68\% & 5.44\% & 70.00\% & -0.03\% & 1.16\% & 1.73\% & 70.00\% & 0.01\% & 0.37\% & 0.56\% \\
 & GMM Estimator & 70.00\% & -1.66\% & 6.47\% & 9.15\% & 70.00\% & -0.53\% & 2.81\% & 4.06\% & 70.00\% & -0.20\% & 1.09\% & 1.60\% \\
\cline{1-14}
\multirow[t]{2}{*}{Partially Observed Network} & Naive Estimator & 70.00\% & -7.11\% & 7.11\% & 4.75\% & 70.00\% & -6.96\% & 6.96\% & 1.43\% & 70.00\% & -6.94\% & 6.94\% & 0.43\% \\
 & GMM Estimator & 70.00\% & -4.09\% & 7.26\% & 9.12\% & 70.00\% & 0.26\% & 4.10\% & 6.29\% & 70.00\% & 0.11\% & 3.41\% & 5.75\% \\
\cline{1-14}
\bottomrule
\end{tabular}
}
    \caption{Simulation: Network Size}
    \label{table:size}
    \floatfoot{\footnotesize{\textit{Notes: This table reports simulation results for the naive and GMM estimator when the graph size is small ($100$ nodes), medium ($1,000$ nodes) and large ($10,000$ nodes). I set the ratio of nodes to links at $1:1$ to maintain sparsity. Bias refers to the median bias, and MAE refers to the median absolute error. The standard error (SE) is computed as the interquartile range of the simulated estimates divided by $1.35$. I simulate $1,000$ times.}}}
\end{table}

\subsection{Non-Gaussian Error Distribution}

The simulation exercise in Table \ref{table:alternative_dist} is designed to examine the sensitivity of the GMM estimator when the latent error distribution is not Gaussian. In the left panel, $U_\ell$ is drawn from a standardized Student's t-distribution with $10$ degrees of freedom. The t-distribution provides a modest deviation from the normal distribution by allowing the error term to have heavier tails. When the network is fully observed and the latent error follows a t-distribution, the naive estimator has smaller bias and MAE than the GMM estimator. 

\begin{table}[!h]
    \centering
    \resizebox{\columnwidth}{!}{
    \begin{tabular}{llrrrrrrrr}
\toprule
 &  & \multicolumn{4}{r}{T Distribution (Degrees of Freedom = 10)} & \multicolumn{4}{r}{Extreme Value Distribution (Shape parameter = 1/2)} \\
 &  & True Value & Bias & MAE & SE & True Value & Bias & MAE & SE \\
\cmidrule(lr){3-6} \cmidrule(lr){7-10}
\midrule
\multirow[t]{2}{*}{Fully Observed Network} & Naive Estimator & 70.00\% & 0.01\% & 0.44\% & 0.64\% & 70.00\% & -3.02\% & 3.02\% & 0.42\% \\
 & GMM Estimator & 70.00\% & 0.88\% & 1.52\% & 2.16\% & 70.00\% & 3.36\% & 3.43\% & 1.98\% \\
\cline{1-10}
\multirow[t]{2}{*}{Partially Observed Network} & Naive Estimator & 70.00\% & -7.46\% & 7.46\% & 0.47\% & 70.00\% & -7.29\% & 7.29\% & 0.38\% \\
 & GMM Estimator & 70.00\% & -1.30\% & 3.59\% & 5.22\% & 70.00\% & 0.40\% & 2.30\% & 3.41\% \\
\cline{1-10}
\bottomrule
\end{tabular}
}
    \caption{Simulation: Non-Gaussian Error}
    \label{table:alternative_dist}
    \floatfoot{\footnotesize{\textit{Notes: This table reports simulation results for the naive and GMM estimators when the DGP error term deviates from normality. The number of nodes is fixed at $10{,}000$, with a node-to-link ratio of $1{:}1$ to maintain sparsity. In the left panel, the standardized t-distribution with $10$ degrees of freedom is used. In the right panel, the generalized extreme value distribution with shape parameter equal to $\frac{1}{2}$ is used (here, I follow the parametrization used by SciPy). Bias refers to the median bias, and MAE refers to the median absolute error. The standard error (SE) is computed as the interquartile range of the simulated estimates divided by $1.35$. I simulate $1,000$ times.}}}
\end{table}

So far, both the normal and $t$ distributions considered have been symmetric with mean zero. The right panel tests robustness to nonzero moments using a generalized extreme value distribution with shape parameter $\frac{1}{2}$, whose first four moments are mean 0.23, variance 0.86, skewness –0.63, and kurtosis 0.25. Both estimators exhibit similar biases when the network is fully observed. Under link truncation, however, the GMM estimator performs far better than the naive estimator. Its bias and MAE are much closer to zero than those of the naive estimator, even in the case of t-distribution. This finding is not surprising given that the left-truncated mean-zero t-distribution has a strictly positive mean.

When the network is partially observed, the GMM estimator effectively corrects the substantial downward bias of the naive estimator—across varying levels of sparsity, network size, and error misspecification—albeit with some loss of efficiency.

\section{Application}\label{sec:application}

In this section, I revisit the classical question of estimating the productivity premium (or loss) in academic collaboration. Although collaboration increases productivity, it also incurs coordination costs such as communication difficulty, shirking, free-riding, and clouded credit assignment (\citet{becker_division_1992}; \citet{jones_rise_2021}). There are various approaches to assess team productivity and to study the relationship between co-authorship and productivity. 

The literature reports various estimates, some of which are of opposite signs. Both \citet{hollis_co-authorship_2001} and \citet{ductor_does_2015} construct their own measures of productivity based on factors such as page length and paper quality.\footnote{Their outcome variable is based on \citet{ductor_social_2014} which is primarily taken from the quality index computed by \citet{kodrzycki_new_2006}. For the journals included in the EconLit database but missing from \citet{kodrzycki_new_2006}, \citet{ductor_social_2014} build a predicted index for them.} The former employs an instrumental variable approach to estimate the productivity premium at approximately 70\%, whereas the latter estimates a productivity loss ranging from $-7\%$ to $-20\%$. Applying a model of generalized means, \citet{ahmadpoor_decoding_2019} estimate the distribution of $\lambda$ across fields and find that co-authorship increases paper impact in most fields, with a median premium of $105\%$. Using the sample from \citet{ductor_social_2014}, \citet{bonhomme_teams_2021} estimates the collaboration premium at around $34\%$. More recently, \citet{anderson_collaborative_2022} fit a regression tree on journal impact score, and also find that larger teams are more productive on average.

Because the measurement of paper quality varies substantially in the aforementioned literature, the comparison of the reported estimates is challenging. What sets this analysis apart from previous work is its emphasis on how much of a \textit{difference} it makes when missing links—unpublished research projects—are taken into account. I apply the GMM estimator to academic publication data in which the quality and quantity of unpublished projects are unobserved. In particular, I examine the premium associated with two-author production relative to single-author production. I find that neglecting the missing links results in a substantial downward bias, necessitating a correction.

\subsection{Data}
The main dataset is Microsoft Academic Graph (MAG), a project run by Microsoft Research that uses machine readers to crawl and collect publication records from the Internet. MAG data are exhaustive in that they include articles from (almost) all scientific fields, spanning hundreds of years of publication. To measure the quality of project $\ell$, I use the impact factor published by Web of Science. The impact factor is the ratio between the total number of citations received by a given journal in the current year and the total number of published articles in the two previous years. Since the raw citation count is non-negative, the impact factor is also non-negative by construction. Another popular measure of paper quality is paper-specific citation count. However, since the MAG initiative was discontinued in 2021, articles published around 2020 may not have sufficient time to accumulate citations, relative to older articles. I therefore use the journal-level impact factor, which has also been adopted by previous studies (e.g., \citet{ductor_social_2014} and \citet{anderson_collaborative_2022}). 

I focus on articles published in ``economics journals'' exclusively written by ``economists''. First, I use Web of Science journal classification to identify economics journals. Second, since the GMM estimator relies on the triplet $(Y_{\ell_{ij}}, Y_{\ell_{i}}, Y_{\ell_{j}})$, I focus on economists who have published at least one single-authored and one co-authored article in economics journals. Although academic collaboration in other disciplines is also of interest, the parsimonious production function used here may be less applicable in other fields. In fields such as the physical and medical sciences, the order of the authors' names matters since the first authors are usually the ones who contribute more. In economics, because most journals use alphabetical ordering, the name order does not provide additional information on individual inputs.\footnote{There are few exceptions. For instance, American Economic Association introduced a randomization tool for authors opting to randomize the name order of their coauthored papers. Nevertheless, the randomized name order remains uninformative on the individual inputs.} This is consistent with the linear specification in (\ref{model:baseline}) that imposes symmetry between individual fixed effects. For the same reason, I do not include interdisciplinary collaboration between economists and non-economists. Although not implemented in this application, one could augment the dataset by collecting each researcher's exact graduation year from publicly available CVs or personal webpages. This information could then be used to construct a time-varying measure of academic experience, as discussed in Extension \ref{model:covariates}.

The sample covers 1997–2020, since the Web of Science Journal Citation Report only began in 1997 and MAG project was shut down in 2021. There are 359 economic journals whose impact factors are available during at least one year between 1997 and 2020. Implementing Algorithm \ref{alg:triplet_construction}, described in the Appendix, yields a sample of 15,875 economists, 25,047 co-authored papers, and 50,094 single-authored papers. Authors publish an average of 3.16 papers, highlighting the sparse structure of the publication network.

There are two potential obstacles to identification and estimation. First, a journal’s quality may change over time. To address this concern, I match detailed \textit{journal-year} impact factor index from Web of Science to individual articles based on their publication year. The second challenge pertains to using fixed effects to model individual types that may evolve over time. Specifically, worker types might grow as workers accumulate more work experience. To mitigate this concern, for each triplet, solo articles are paired with co-authored articles published in a similar time period where possible.

\subsection{Collaboration Network with Missing Links}

Before selecting an estimator, I first test the null hypothesis that the academic collaboration network is fully observed. To implement the J-test in Lemma \ref{J_test}, I construct additional valid moments by interacting (\ref{moment:linear_latent}) with network statistics computed from the observed graph $\mathcal{G}$. Among various admissible network statistics, I use node degree and closeness centrality in this application.\footnote{Testing different network statistics is recommended, since some statistics generate little variation at the node or dyadic level, and could lead to false negative results when used as over-identifying moments.} Because these statistics are computed at the node level, I take a simple sum at the dyadic level before interacting them with (\ref{moment:linear_latent}). I then perform a two-step GMM estimation, obtaining a J-test statistic of $6.56$.\footnote{The two-step procedure yields an efficient GMM estimator by using the estimated optimal weighting matrix when there are $K$ over-identifying restrictions.} In particular, the asymptotic distribution of the test statistics is chi-squared with two degrees of freedom, since the number of parameters is one (i.e. $\lambda$) and the number of moments is three. This yields a $p$-value of $0.0376$. At the $5\%$ significance level, I reject the null hypothesis that the collaboration network is fully observed.

\subsection{Correcting Underestimated Collaboration Premium}\label{sec:premium_correction}

The J-test implemented in the previous section suggests the need to account for missing links. Assuming full network observability, the naive estimator is misspecified and gives $\hat\lambda_{\text{naive}} = 0.584$ with $90\%$ bootstrap CI of $[0.580, 0.590]$. To address plausible truncation bias, I employ the GMM estimator from Theorem \ref{theorem:gmm}. I find that $\hat\lambda$ rises to $0.651$—a gain of over 10\%—with a standard error of $0.04$ and 90\% CI of $[0.584, 0.718]$. Although the difference between the estimated $\lambda$ may appear small in absolute terms, it is substantial given that the scaling factor is typically bounded between 0 and 1.

\begin{figure}[htbp]
    \centering
    \includegraphics[width=0.6\linewidth]{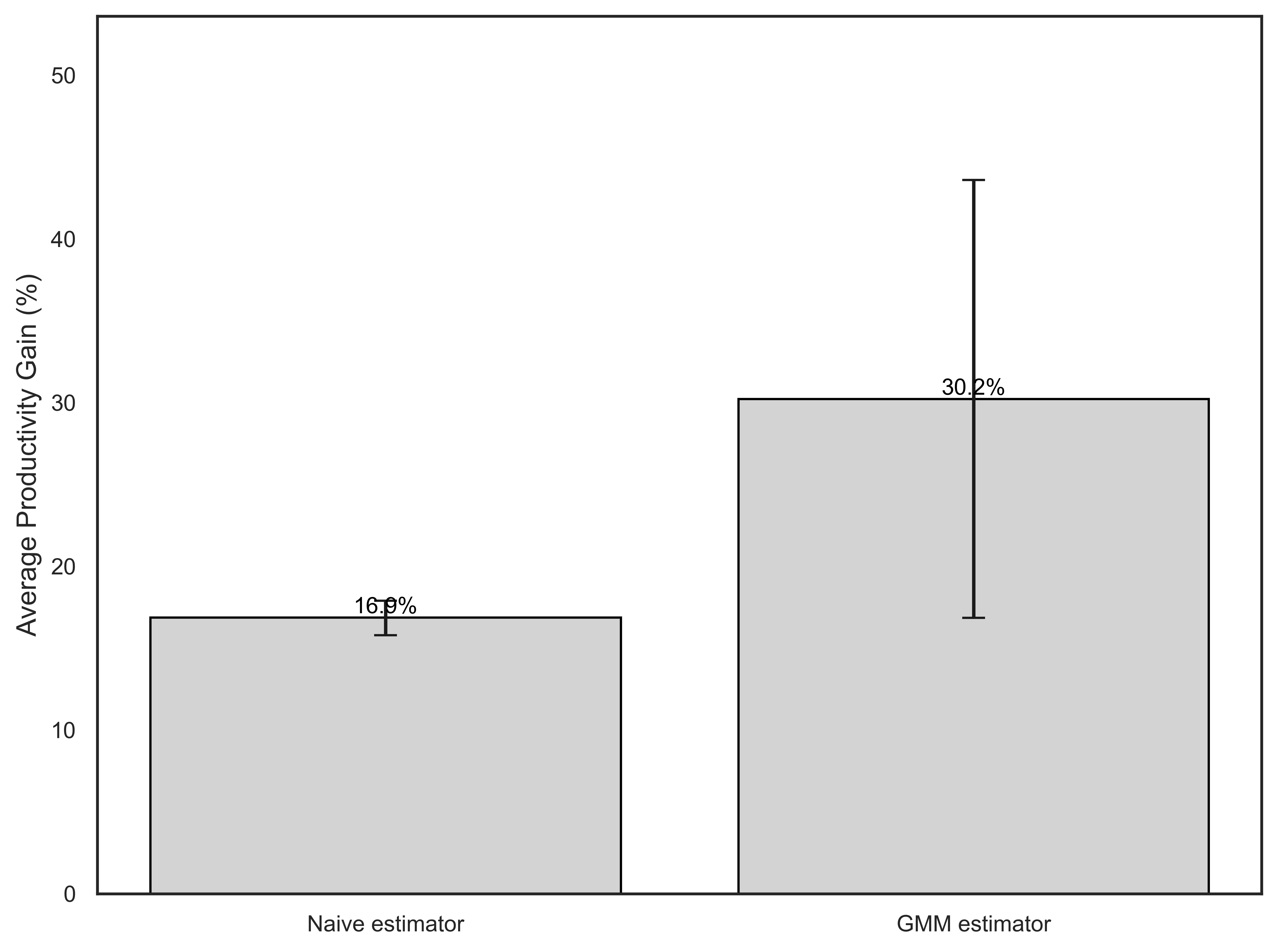}
    \caption{Average Productivity Gain}
    \label{fig:average_productivity_gain}
    \floatfoot{\footnotesize{\textit{Notes: This bar plot shows the average collaboration premium implied by the naive estimate and the GMM estimates, with $90\%$ confidence intervals. The outcome variable is measured by the journal impact factor.}}}
\end{figure}

To better interpret the two estimates, I compute the average productivity gain implied by the team-size scaling factor. Suppose that two researchers of identical type work together, the average productivity gain (computed as $\hat\lambda \cdot 2 - 1$) is $30.2\%$. This implies an almost 30 percent increase in the paper quality (Figure \ref{fig:average_productivity_gain}), doubling the estimated premium from the naive estimator.
The downward truncation bias suffered by $\hat\lambda_{\text{naive}}$ is not surprising. The left-sided truncation at zero disproportionately favors ``lucky'' projects that receive large and positive shocks. This artificially inflates the workers' input and hence deflates the estimated scaling factor.

So far, I have used the journal impact factor index to assess paper quality. For robustness checks, I re-estimate the model using the Clarivate's Eigenfactor Metrics (see \citet{west_closer_2017}). The eigenfactor index measures paper quality by tracking citations from high-impact articles, using the directed citation network to gauge journal influence and, by extension, article quality of a particular journal. The naive and the GMM estimates are respectively $0.586$ with 90\% CI of $[0.576, 0.597]$, and $0.670$, with 90\% CI of $[0.569, 0.770]$. The results obtained from both metrics are consistent (see Appendix Figure \ref{fig:average_productivity_gain_eigenfactor}).

\subsection{Evidence for Time-varying Productivity Gain}

Another natural question is whether the collaboration premium varies over time—potentially influenced by advances in information technology (IT) that enhance teamwork productivity (see \citet{dulebohn_virtual_2017} and \citet{karl_virtual_2022}). Prior to the launch of video conference tools, notably Skype in 2003, email was the primary communication channel for geographically dispersed researchers. 
\begin{figure}[!ht]
    \centering
    \includegraphics[width=0.6\linewidth]{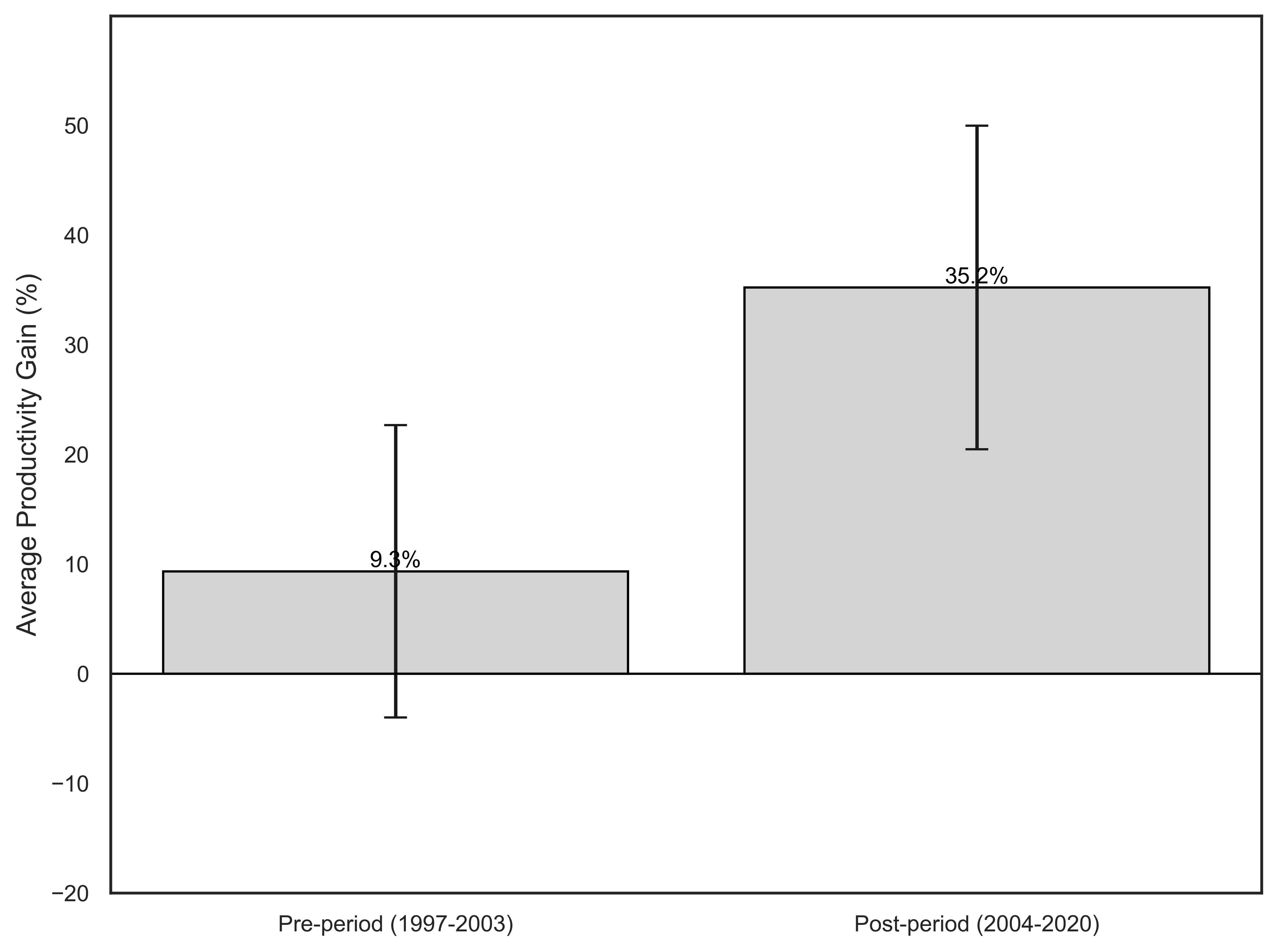}
    \caption{Time-varying Productivity Gain}
    \label{fig:pre_post_productivity_gain}
    \floatfoot{\footnotesize{\textit{Notes: This bar plot shows the average collaboration premium estimated by the GMM estimator in the pre-2003 and post-2003 periods, with $90\%$ confidence intervals. The outcome variable is measured by the journal impact factor.}}}
\end{figure}
Motivated by the introduction of video conference tools in early 2000s, I divide the sample into two periods: 1997–2003 and 2004-2020. However, I do not make causal claims about the effect of IT, given potential omitted variables and the absence of data on individual-specific timing of IT adoption. I apply the GMM estimator for each time period to separately estimate the average collaboration premiums. 

Figure \ref{fig:pre_post_productivity_gain} reveals a stark difference between the two estimated premiums: the average productivity gain of $35.2\%$ after 2003 appears significantly higher than the $9.3\%$ gain for the 1997--2003 sample. It also exceeds the $30.2\%$ gain from the pooled 1997--2020 sample. The pre-2003 confidence interval covers zero, while the post-2003 estimate is statistically significant—suggesting an increase in the collaboration premium over time.

\section{Conclusion}\label{sec:conclusion}
This paper studies a team production model with missing links. Researchers often face data constraints that prevent full network observation. In partially observed networks, both links and nodes may be selectively omitted, which can introduce significant bias that requires correction. Because partial network observability correlates the observed graph with unobserved errors, standard methods that assume a fully observed exogenous network may fail. Despite the importance of accounting for missing links and nodes, research on models with partial network observability is still limited. This paper proposes a truncation-robust GMM estimator and a test for detecting link truncation, and shows empirically that unaccounted missing links systematically bias estimates of the collaboration premium. While this paper assumes a common truncation cutoff, future work could extend the model to allow heterogeneous individual thresholds. Other promising directions include relaxing the normality assumption on the error distribution and developing consistent estimators of individual fixed effects in partially observed networks.

\section*{Acknowledgements}
I am extremely grateful for the generous guidance of Martin Weidner. I also thank Isaiah Andrews, Helmut Farbmacher, Raffaella Giacomini, Peter Hull, Martin Mugnier, Taisuke Otsu, Xiaoxia Shi, and Frank Windmeijer for helpful discussion and comments. I am responsible for all errors.

\newpage
\section*{References}
\renewcommand{\refname}{}
\bibliography{truncated}

\section*{Appendix A: Proofs of Results}
\textbf{Proof of Lemma \ref{lemma:TruncatedNormal}:} Denote the $k^{\text{th}}$ moment of the truncated normal $Y$ as
$
\mu_k := \EE[\widetilde{Y}^k | \widetilde{Y} \ge 0],
$
where $\mu_0 = 1$. Let $\phi$ and $\Phi$ denote the PDF and CDF of the standard normal distribution. Then,
\begin{align*}
    \mu_{k+1} 
    &= \int_0^\infty y^{k+1} \cdot\left(\frac{1}{\tilde{\sigma}} \frac{\phi(\frac{y-\tilde{\alpha}}{\tilde{\sigma}})}{1 - \Phi(\frac{-\tilde{\alpha}}{\tilde{\sigma}})} \right) \,dy \\
    &= \frac{1}{1 - \Phi(\frac{-\tilde{\alpha}}{\tilde{\sigma}})} \int_0^\infty y^{k} \left(\frac{y-\tilde{\alpha}}{\tilde{\sigma}^2} + \frac{\tilde{\alpha}}{\tilde{\sigma}^2}\right) \cdot\tilde{\sigma}^2 \left(\frac{1}{\tilde{\sigma}} \phi\left(\frac{y-\tilde{\alpha}}{\tilde{\sigma}}\right) \right) \,dy \\
    &= \tilde{\sigma}\frac{1}{1 - \Phi(\frac{-\tilde{\alpha}}{\tilde{\sigma}})} \int_0^\infty y^{k} \left(\frac{y-\tilde{\alpha}}{\tilde{\sigma}^2} \cdot \phi\left(\frac{y-\tilde{\alpha}}{\tilde{\sigma}}\right)\right) \,dy +
    \tilde{\alpha} \left[\frac{1}{1 - \Phi(\frac{-\tilde{\alpha}}{\tilde{\sigma}})} \int_0^\infty y^{k} \cdot\left(\frac{1}{\tilde{\sigma}} \phi\left(\frac{y-\tilde{\alpha}}{\tilde{\sigma}}\right) \right) \,dy\right]
\end{align*}
Note the second term is simply $\mu_{k}$. Recall that $\phi(z)' = (-z)\phi(z)$. Applying integration by parts to the first term,
\begin{align*}
    \mu_{k+1} 
    &= \tilde{\sigma} \frac{-1}{1 - \Phi(\frac{-\tilde{\alpha}}{\tilde{\sigma}})} \int_0^\infty y^{k} \left(-\frac{y-\tilde{\alpha}}{\tilde{\sigma}} \cdot \frac{1}{\tilde{\sigma}} \phi\left(\frac{y-\tilde{\alpha}}{\tilde{\sigma}}\right)\right) \,dy + \tilde{\alpha} \mu_k \\
    &= - \tilde{\sigma} \left[\frac{1}{1 - \Phi(\frac{-\tilde{\alpha}}{\tilde{\sigma}})} \int_0^\infty y^{k} \left(\frac{d}{d y} \phi\left(\frac{y-\tilde{\alpha}}{\tilde{\sigma}}\right)\right) \,dy\right] + \tilde{\alpha} \mu_k \\
    &= \tilde{\alpha} \mu_k  - \frac{\tilde{\sigma}}{1 - \Phi(\frac{-\tilde{\alpha}}{\tilde{\sigma}})} \left[y^k \phi\left(\frac{y-\tilde{\alpha}}{\tilde{\sigma}}\right) \bigg|_0^\infty - \int_0^\infty (k y^{k-1}) \cdot \phi\left(\frac{y-\tilde{\alpha}}{\tilde{\sigma}}\right) \,dy  \right] \\
    &= \tilde{\alpha} \mu_k  + (k\tilde{\sigma}) \tilde{\sigma} \left(\int_0^\infty y^{k-1} \frac{1}{\tilde{\sigma}} \frac{\phi\left(\frac{y-\tilde{\alpha}}{\tilde{\sigma}}\right)}{1 - \Phi(\frac{-\tilde{\alpha}}{\tilde{\sigma}})} \,dy \right) \\
    &= \tilde{\alpha} \mu_k  + k \tilde{\sigma}^2 \mu_{k-1}.
\end{align*}
Therefore,
$
\mu_{k+1} = \widetilde{\alpha} \mu_{k} + k \widetilde{\sigma}^2 \mu_{k-1}
$.
After rearranging,
\begin{align*}
    \EE[\widetilde{Y}^{k+1} - \widetilde{\alpha} \widetilde{Y}^k - k \widetilde{\sigma}^2 \widetilde{Y}^{k-1} | \widetilde{Y} \ge 0] = 0. \tag*{\text{\(\Box\)}}
\end{align*}

\medskip
\textbf{Proof of Proposition \ref{prop:moments}:} First, I apply Lemma \ref{lemma:TruncatedNormal} three times to $Y_{\ell_{i}} \sim N(\alpha_i, \sigma^2)$, $Y_{\ell_{j}} \sim N(\alpha_j, \sigma^2)$, and $Y_{\ell_{ij}} \sim N(\lambda(\alpha_i + \alpha_j), \sigma^2)$, respectively. This gives the following system of equalities:
\begin{align*}
\EE[Y_{\ell_{i}}^{k+1} - \alpha_i Y_{\ell_{i}}^k - k \sigma^2 Y_{\ell_{i}}^{k-1} | Y_{\ell_{i}}^* \ge 0] = 0,
\\
\EE[Y_{\ell_{j}}^{k+1} - \alpha_j Y_{\ell_{j}}^k - k \sigma^2 Y_{\ell_{j}}^{k-1} | Y_{\ell_{i}}^* \ge 0] = 0,
\\
\EE[Y_{\ell_{ij}}^{k+1} - \lambda(\alpha_i + \alpha_j)Y_{\ell_{ij}}^k - k\sigma^2 Y_{\ell_{ij}}^{k-1} | Y_{\ell_{ij}}^* \ge 0] = 0.    
\end{align*}
By Assumption (\ref{assumption:iid}), the above moments also hold conditional on the fixed effects.
\begin{align*}
\EE[Y_{\ell_{i}}^{k+1} - \alpha_i Y_{\ell_{i}}^k - k \sigma^2 Y_{\ell_{i}}^{k-1} | Y_{\ell_{i}}^* \ge 0, Y_{\ell_{j}}^* \ge 0, Y_{\ell_{ij}}^* \ge 0, \alpha_i, \alpha_j] = 0,
\\
\EE[Y_{\ell_{j}}^{k+1} - \alpha_j Y_{\ell_{j}}^k - k \sigma^2 Y_{\ell_{j}}^{k-1} | Y_{\ell_{i}}^* \ge 0, Y_{\ell_{j}}^* \ge 0, Y_{\ell_{ij}}^* \ge 0, \alpha_i, \alpha_j] = 0,
\\
\EE[Y_{\ell_{ij}}^{k+1} - \lambda(\alpha_i + \alpha_j)Y_{\ell_{ij}}^k - k\sigma^2 Y_{\ell_{ij}}^{k-1} | Y_{\ell_{i}}^* \ge 0, Y_{\ell_{j}}^* \ge 0, Y_{\ell_{ij}}^* \ge 0, \alpha_i, \alpha_j] = 0.
\end{align*}
First, scale each equation by $\lambda$. Next, multiply the first equation by $Y_{\ell_j}^k Y_{\ell_{ij}}^k$, the second by $Y_{\ell_i}^k Y_{\ell_{ij}}^k$, and the third by $Y_{\ell_i}^k Y_{\ell_j}^k$. These transformations preserve equality because the shocks are i.i.d.\ by Assumption~(\ref{assumption:iid}). Let $\EE^\dagger$ denote conditional expectation given $Y_{\ell_i}^*, Y_{\ell_j}^*, Y_{\ell_{ij}}^* \ge 0$. Then,
\begin{align*}
\EE^\dagger[\lambda Y_{\ell_{i}}^{k+1} Y_{\ell_{j}}^k Y_{\ell_{ij}}^k - \lambda \alpha_i Y_{\ell_i}^k Y_{\ell_j}^k Y_{\ell_{ij}}^k - k \lambda \sigma^2 Y_{\ell_{i}}^{k-1} Y_{\ell_{j}}^k Y_{\ell_{ij}}^k | \alpha_i, \alpha_j] = 0,
\\
\EE^\dagger[\lambda Y_{\ell_{i}}^k Y_{\ell_{j}}^{k+1} Y_{\ell_{ij}}^k - \lambda \alpha_j Y_{\ell_i}^k Y_{\ell_j}^k Y_{\ell_{ij}}^k - k \lambda \sigma^2 Y_{\ell_{i}}^k Y_{\ell_{j}}^{k-1} Y_{\ell_{ij}}^k | \alpha_i, \alpha_j] = 0,
\\
\EE^\dagger[Y_{\ell_{i}}^k Y_{\ell_{j}}^k Y_{\ell_{ij}}^{k+1} - \lambda(\alpha_i + \alpha_j) Y_{\ell_i}^k Y_{\ell_j}^k Y_{\ell_{ij}}^k  - k\sigma^2 Y_{\ell_{i}}^k Y_{\ell_{j}}^k Y_{\ell_{ij}}^{k-1} | \alpha_i, \alpha_j] = 0.  
\end{align*}
Next, difference out $(\alpha_i, \alpha_j)$ by subtracting the sum of the first two equations from the third. This gives
$$
\EE^\dagger[(Y_{\ell_{i}}^k Y_{\ell_{j}}^k Y_{\ell_{ij}}^{k+1} - \lambda Y_{\ell_{i}}^{k+1} Y_{\ell_{j}}^k Y_{\ell_{ij}}^k - \lambda Y_{\ell_{i}}^k Y_{\ell_{j}}^{k+1} Y_{\ell_{ij}}^k) +  k\sigma^2(\lambda Y_{\ell_{i}}^{k-1} Y_{\ell_{j}}^k Y_{\ell_{ij}}^k + \lambda Y_{\ell_{i}}^k Y_{\ell_{j}}^{k-1} Y_{\ell_{ij}}^k - Y_{\ell_{i}}^k Y_{\ell_{j}}^k Y_{\ell_{ij}}^{k-1}) | \alpha_i, \alpha_j] = 0.
$$
By the law of iterated expectations,
$$
\EE^\dagger[(Y_{\ell_{i}}^k Y_{\ell_{j}}^k Y_{\ell_{ij}}^{k+1} - \lambda Y_{\ell_{i}}^{k+1} Y_{\ell_{j}}^k Y_{\ell_{ij}}^k - \lambda Y_{\ell_{i}}^k Y_{\ell_{j}}^{k+1} Y_{\ell_{ij}}^k) +  k\sigma^2(\lambda Y_{\ell_{i}}^{k-1} Y_{\ell_{j}}^k Y_{\ell_{ij}}^k + \lambda Y_{\ell_{i}}^k Y_{\ell_{j}}^{k-1} Y_{\ell_{ij}}^k - Y_{\ell_{i}}^k Y_{\ell_{j}}^k Y_{\ell_{ij}}^{k-1})] = 0.
$$
After rearranging,
\begin{align*}
\EE^\dagger[Y_{\ell_i}^k Y_{\ell_j}^k Y_{\ell_{ij}}^k (Y_{\ell_{ij}} - \lambda(Y_{\ell_i} + Y_{\ell_j})) + k\sigma^2  Y_{\ell_i}^{k-1} Y_{\ell_j}^{k-1} Y_{\ell_{ij}}^{k-1} (\lambda(Y_{\ell_i} + Y_{\ell_j})Y_{\ell_{ij}}  -  Y_{\ell_{i}} Y_{\ell_{j}})] = 0. \tag*{\text{\(\Box\)}}
\end{align*}

\medskip
\textbf{Proof of Theorem \ref{theorem:gmm}:} I verify the conditions in \citet{hansen_large_1982}. First, I check that the limiting gradient vector 
$$
G = \EE^\dagger \begin{bmatrix}
\frac{\partial m(\lambda, \sigma, Y_{\ell_i}, Y_{\ell_j}, Y_{\ell_{ij}})}{\partial \lambda}
\frac{\partial m(\lambda, \sigma, Y_{\ell_i}, Y_{\ell_j}, Y_{\ell_{ij}})}{\partial \sigma}
\end{bmatrix}
$$
exists. Second, I verify that the sample variance of the moments converges to 
$$
V = \Var^\dagger(m(\lambda, \sigma^2, Y_{\ell_i}, Y_{\ell_j}, Y_{\ell_{ij}})).
$$ 
Finally, as noted in Remark \ref{remark:rank}, it suffices to verify the rank condition of the linear system of moments for identification. If these conditions hold, the GMM estimator is asymptotically distributed as
\begin{equation*}
    \sqrt{|B|}
        \left[
        \begin{pmatrix}
        \hat\lambda \\ \hat\sigma^2
        \end{pmatrix} 
        -
        \begin{pmatrix}
        \lambda \\ \sigma^2
        \end{pmatrix}
        \right]
        \Longrightarrow
        N\left(0, (G' V^{-1} G)^{-1}\right). \tag*{\text{\(\Box\)}}
\end{equation*}

\newpage
\section*{Appendix B: Monte Carlo Simulation}
\begin{algorithm*}[!ht]
    \caption{Simulation} \label{alg:simulation}
    \begin{enumerate}
        \item \textbf{Latent Types:}\\
        Draw $N$ individual fixed effects $\{\alpha_i\}_{i=1}^N \simiid \mathrm{Pareto}(0,10)$.
        \item \textbf{Solo Production:}\\
        Individual $i$ produces a solo project of latent quality $Y_{\ell}^*$ as in (\ref{model:baseline}).
        \item \textbf{Team Formation:}\\
        Randomly match pairs of individuals; control network sparsity via the target average degree (average number of links per node).
        \item \textbf{Team Production:}\\
        For each team project $\ell$, its latent quality $Y_{\ell}^*$ follows (\ref{model:baseline}). 
        \item \textbf{Link Truncation:}\\
        Observe project $\ell$ if and only if $Y_{\ell}^* \ge 0$. In that case, set the observed outcome $Y_{\ell} = Y_{\ell}^*$; otherwise the project (and link) is unobserved.
        \item \textbf{Triplet Construction:}\\
        Construct triplets $(Y_{\ell_i}, Y_{\ell_j}, Y_{\ell_{ij}})$ such that each project appears in at most one triplet, which ensures independence across triplets.
        \item \textbf{Estimation:}\\
        Using both (i) the latent network and (ii) the observed (truncated) network,
        \begin{enumerate}
            \item compute the naive estimator; 
            \item compute the truncation-robust GMM estimator.
        \end{enumerate}
    \end{enumerate}
\end{algorithm*}

\FloatBarrier
\section*{Appendix C: Triplet Construction}
A \textit{triplet} is the tuple $(Y_{\ell_{ij}}, Y_{\ell_i}, Y_{\ell_j})$ linking the team project $Y_{\ell_{ij}}$ of workers $i$ and $j$ to their solo projects $(Y_{\ell_i}, Y_{\ell_j})$. The asterisk is omitted since truncated links are unobserved. Because collaboration patterns vary widely, a systematic procedure is required to construct triplets: some workers have many team projects but few solo ones (or vice versa), creating multiple possible matchings. The algorithm below (i) ensures independence across triplets by avoiding project reuse and (ii) uses timestamps (e.g., publication dates) to optimally pair solo and team projects, mitigating time variation in individual types.
\begin{algorithm*}[!ht]
    \caption{Construction of Independent Triplets}
    \medskip
    \label{alg:triplet_construction}
    {\raggedright Iterate until all team projects are matched or dropped.\par}
    \begin{enumerate}
        \item For each team project $\ell$ with outcome $Y_{\ell_{ij}}$, identify its two workers $i$ and $j$.
        \item List all unmatched solo projects for $i$ and $j$.
        \item Drop team project $\ell$ if either worker has no unmatched solo project.
        \item Otherwise, for each worker:
            \begin{itemize}
                \item If only one solo project exists, match it to $Y_{\ell_{ij}}$.
                \item If multiple, match the solo project closest in time to $Y_{\ell_{ij}}$.
            \end{itemize}
        \item Record the triplet $(Y_{\ell_i}, Y_{\ell_j}, Y_{\ell_{ij}})$ and remove matched projects.
    \end{enumerate}
\end{algorithm*}

\textbf{Example:} Suppose $A$ and $B$ co-authored a paper in 2010. $A$ has solo papers in 2006 and 2011, and $B$ has one in 2015. The algorithm matches $A$'s 2011 and $B$'s 2015 papers to their joint 2010 paper.

\bigskip
\section*{Appendix D: Robustness check: Eigenfactor Index}
\begin{figure}[H]
    \centering
    \includegraphics[width=0.6\linewidth]{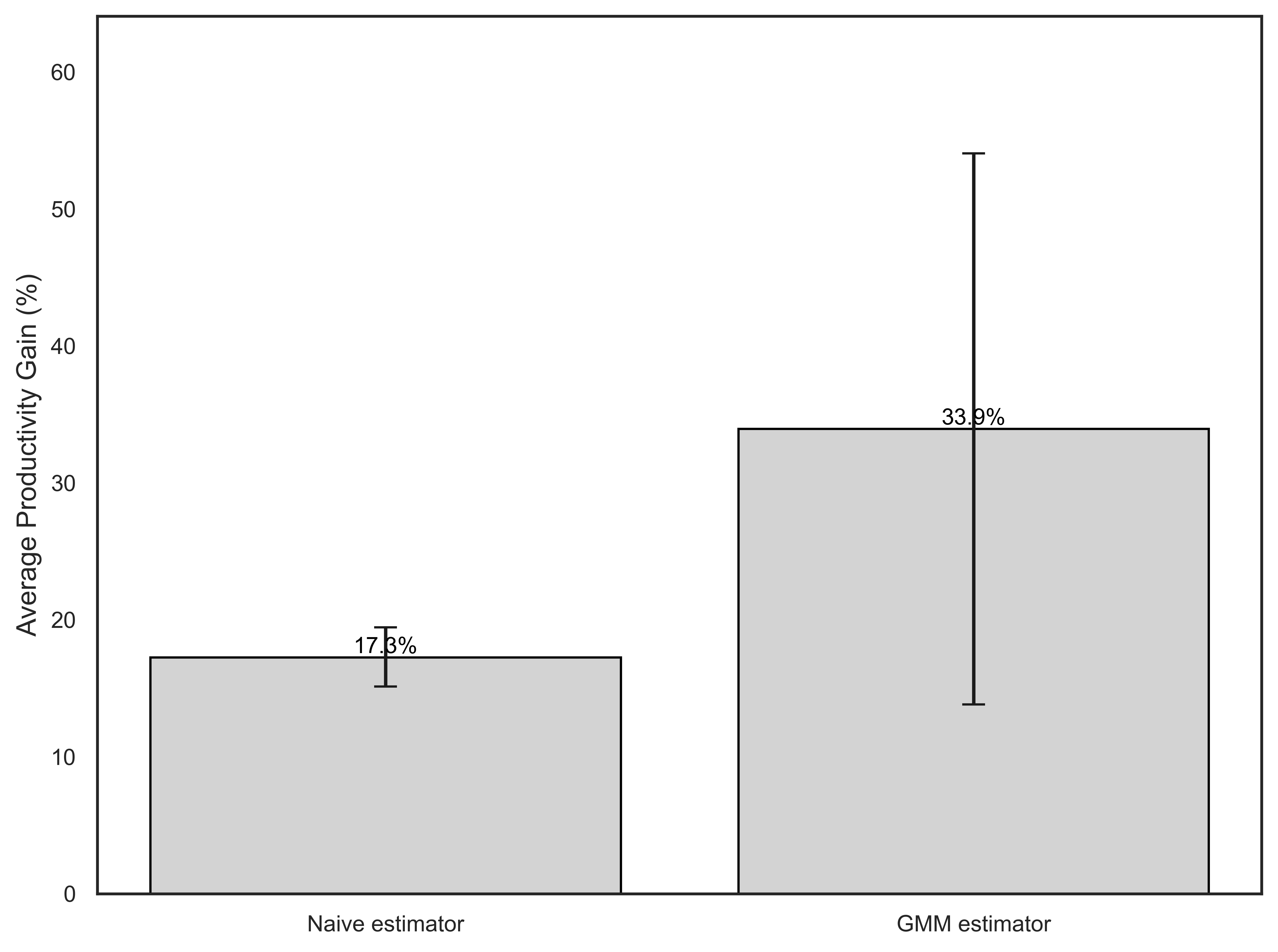}
    \caption{Average Productivity Gain (Eigenfactor Index)}
    \label{fig:average_productivity_gain_eigenfactor}
    \floatfoot{\footnotesize{\textit{Notes: This bar plot shows the average collaboration premium implied by both the naive estimate and the GMM estimates, with $90\%$ confidence intervals. The outcome variable is measured by the Eigenfactor index.}}}
\end{figure}

\end{document}